\documentclass[aps,pra,superscriptaddress,10pt,a4paper,showkeys,twocolumn]{revtex4-2}
\usepackage{amsmath,amssymb,amsthm,mathtools,bm}
\usepackage{natbib}
\usepackage{graphicx}
\usepackage[hidelinks]{hyperref}
\usepackage[capitalise,nameinlink]{cleveref}
\usepackage{overpic}
\usepackage{siunitx}
\usepackage{xcolor}
\usepackage{xfrac}
\usepackage{comment}
\usepackage[normalem]{ulem}

\renewcommand{\vec}[1]{\bm{#1}}
\newcommand{\diff}[2]{\frac{\mathrm{d}#1}{\mathrm{d}#2}}
\newcommand{\abs}[1]{\left\lvert#1\right\rvert}
\newcommand{\avg}[1]{\langle#1\rangle}
\newcommand{\intd}[1]{\,\mathrm{d}#1}
\newcommand{\phaseshift}{\phi}

\begin{document} 
\title{Tunable asymmetric swimming in biflagellate microswimmers}

\author{Benjamin J. Walker} 
\email{benjamin.walker@ucl.ac.uk}
\affiliation{Department of Mathematics, University College London, London, WC1E 6BT, UK}

\author{Cl\'ement Moreau} 
\email{clement.moreau@cnrs.fr}
\affiliation{Nantes Universit\'e, \'Ecole Centrale Nantes, CNRS, LS2N, UMR 6004, F-44000 Nantes, France}

\author{Tommie L. Robinson} 
\affiliation{ School of Physics, Georgia Institute of Technology, Atlanta, GA 30332, United States of America}

\author{Zhaochen J. Xu} 
\affiliation{ School of Physics, Georgia Institute of Technology, Atlanta, GA 30332, United States of America}

\author{Daniel I. Goldman} 
\affiliation{ School of Physics, Georgia Institute of Technology, Atlanta, GA 30332, United States of America}

\author{Eamonn A. Gaffney} 
\email{gaffney@maths.ox.ac.uk}
\affiliation{Mathematical Institute, University of Oxford, Oxford, OX2 6GG, United Kingdom}

\author{Kirsty Y. Wan} 
\email{k.y.wan2@exeter.ac.uk}
\affiliation{Living Systems Institute \& Department of Mathematics and Statistics, University of Exeter, Exeter EX4 4QD, United Kingdom}
\date{\today}

\begin{abstract}
\color{black}
Many biological microswimmers can modulate their swimming gait to achieve directional control of motility, especially when performing steering towards specific directional cues. This can be achieved without the need for obvious morphological or structural asymmetries in the form of the organism, or in the number or organisation of propulsion-generating appendages such as cilia. In this work, we identify and validate a core principle of asymmetric turning in biflagellate microswimmers: propulsive forces interact constructively to drive translation whilst interacting destructively to drive rotation. We explore the ramifications of this tunable biflagellar swimming mechanism across a range of systems, from a simple, back-of-the-envelope model to a detailed computational representation of an exemplar swimmer. This leads to a markedly general quantitative relation between key drivers of asymmetry, such as ciliary beat frequency, and the curvature of emergent trajectories. We discuss how the model green alga \textit{Chlamydomonas reinhardtii}, which actuates its two cilia in a symmetric breaststroke for forward swimming, may exploit this feature for phototaxis. Finally, we validate our predictions in a \textit{Chlamydomonas}-inspired robophysical model, implementing closed-loop control to achieve phototactic turning.
\end{abstract}

\maketitle

\section*{Introduction}

\begin{figure}
    \centering
    \includegraphics[width=\linewidth]{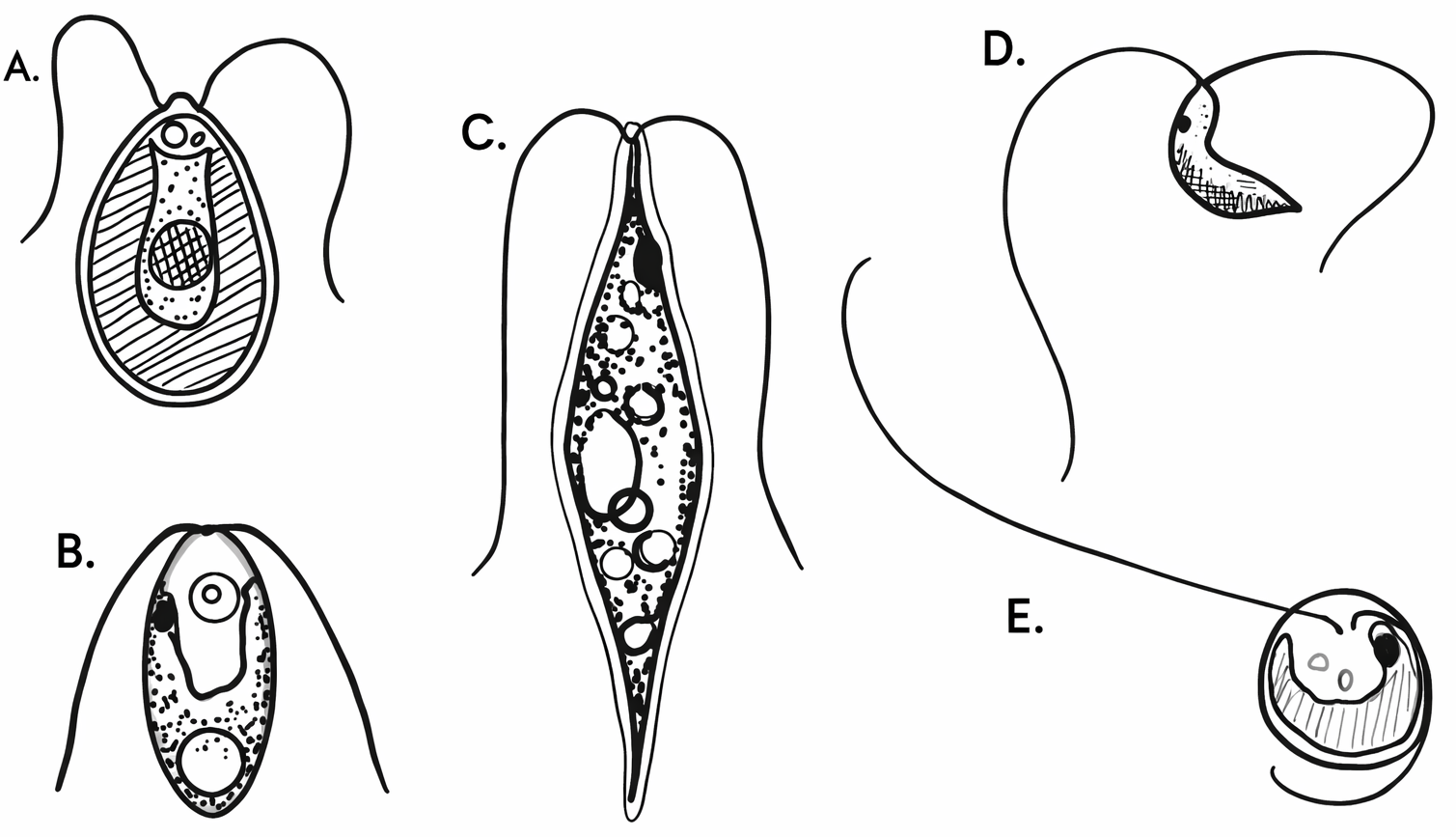}
    \caption{Examples of common biflagellate microswimmers found in nature (sizes range from \qtyrange{5}{20}{\micro\metre}). Illustrated are morphologies representative of each genus. A. \textit{Chlamydomonas}, B. \textit{Dunaliella}, C. \textit{Chlorogonium}, D. \textit{Spermatozopsis}, E. \textit{Nephroselmis}.}
    \label{fig: bio examples}
\end{figure}

Achieving directional control of motility is a fundamental challenge of swimming at low-Reynolds number \cite{berg1977physics,purcell2014life}. For small prokaryotic organisms of characteristic size \qtyrange[range-units=single,range-phrase=-]{1}{2}{\micro\metre}, swimming trajectories are easily overwhelmed by rotational diffusion; consequently, few such organisms have evolved the capacity to steer \cite{Wan2021}. In contrast, larger eukaryotic cells perform more elaborate strategies for controlling their heading in a stimulus-dependent manner, often involving the coordinated movement of multiple cilia (also known as eukaryotic flagella). Despite the diversity of naturally occurring motility and navigation mechanisms, actively migrating cells (such as algae, diatoms, amoeba, immune cells and cancer cells) universally assume complex trajectories with time-varying curvatures. This suggests that the ability to adjust motility in response to internal or external cues is a fundamental phenomenon across species \cite{tokarova2021patterns,ishikawa2025physics,lo2000cell}.

In many cases, the swimmer itself exhibits some persistent geometrical asymmetry, which gives rise to symmetry breaking. 
Uniflagellates including spermatozoa rely on beat waveform modulation to reorient themselves \cite{saggiorato2017human}, whereas biflagellate zoospores with heterodynamic flagella undergo rapid turning events by alternating the activity of the anterior and posterior flagella \cite{tran2022coordination}. 
A similar process occurs in cirri-bearing hypotrich ciliates, which undergo specialised manoeuvres for turning (called side-stepping reactions) \cite{laeverenz2024bioelectric,lueken1996rhythmic}.
Other cells have no apparent morphological asymmetries, such as the biflagellate green alga \textit{Chlamydomonas} shown in \cref{fig: bio examples}, but can nonetheless fine tune the actuation of their two symmetrically positioned, equal-length flagella.
While individual studies have attempted to explore the coupling between swimmer geometry, actuation and motility trajectories \cite{bennett2015steering,hokmabad2021emergence,kummel2013circular}, investigation of how asymmetric drivers of motility relate to the asymmetry or curvature of the resulting trajectories is lacking. %
Comprehensive theoretical understanding of how asymmetries in propulsion are propagated has important consequences not only for decoding the hidden biochemical processes regulating cell motility, but also for the design of novel synthetic swimmers.

Owing to the marked complexity of microscale swimming, the most faithful theoretical models of cell-scale swimmers require intricate descriptions of the geometry of the swimmer and high resolution hydrodynamics to capture interactions with the fluid environment. 
This model specificity can afford significant advantages, such as enabling more precise prediction of swimmer behaviours that accounts for various confounding factors, such as the presence of fluid boundaries or other swimmers \cite{elgeti2010hydrodynamics,smith2009human,fuchter2023three}. %
On the other hand, simple models are often better suited to generalisation, with proven potential to generate broad insight into fundamental principles of cellular swimming \cite{contino2015microalgae,cortese2021control,friedrich2012flagellar}.
It is worth highlighting that, whilst the simplicity and often wide-reaching applicability of minimal models can be beguiling, their use is predicated on appropriate, model-specific caveats and acknowledgement of the constraints on their validity.

In this study, we establish a general, quantitative relationship between cell-scale asymmetries in propulsion and the resulting swimming trajectories. To this end, we analyse simple models of a biflagellate swimmer that comprise a body with two periodically driven appendages, abstracting away the details of ciliary waveforms and, to some extent, swimmer geometry.
We develop and investigate a sequence of theoretical models of increasing complexity: a minimal toy model that accounts for essential features of our exemplar swimmer; an improved model that better captures relevant physics; and a geometrically faithful computational model. 
In each case, we focus on the impact of measurable physical parameters (such as beat amplitude, frequency and phase difference) on the resulting dynamics, and establish conditions for the validity of our predictions. We complement these theoretical insights by realising a robophysical model of biflagellar swimming that, in contrast, requires no specific assumptions on the detailed nature of the fluid-structure interactions involved. 
Finally, we implement proof-of-concept closed-loop phototaxis in our robotic model, motivated by living organisms' routine leveraging of tunable asymmetries in motility to produce deterministic steering in response to environmental cues \cite{tsang2024light,wang2025light,leptos2023phototaxis,xiong2024light}. These robophysical explorations validate the predictions of our theoretical modelling in practice.

\section{Minimal model}\label{sec: minimal model}
We first pose a general, toy model for planar cilia-driven motion. We focus on what will prove to be a defining feature of the dynamics: the constructive interaction of two cilia-like appendages when generating propulsion and their contrastingly destructive interactions when generating rotation. In other words, we explore the consequences of the simple observation that it is the \emph{sum} of the forces generated by the appendages that yields translation, whilst it is their \emph{difference} that produces rotation.

\begin{figure*}
    \centering
    \begin{overpic}[width=0.32\textwidth]{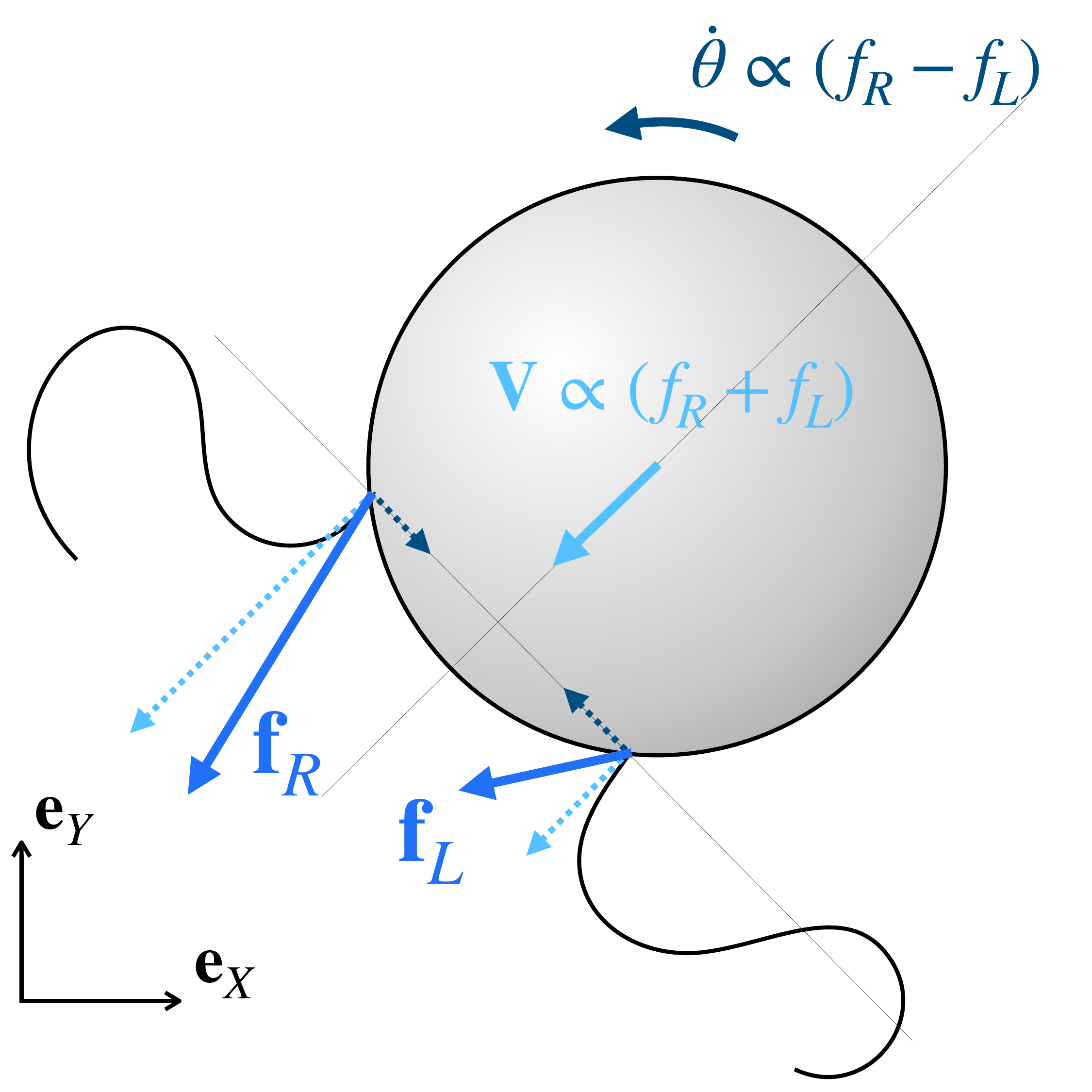}
        \put(0,100){(a)}
    \end{overpic}
    \hspace{0.15\textwidth}
    \begin{overpic}[width=0.32\textwidth]{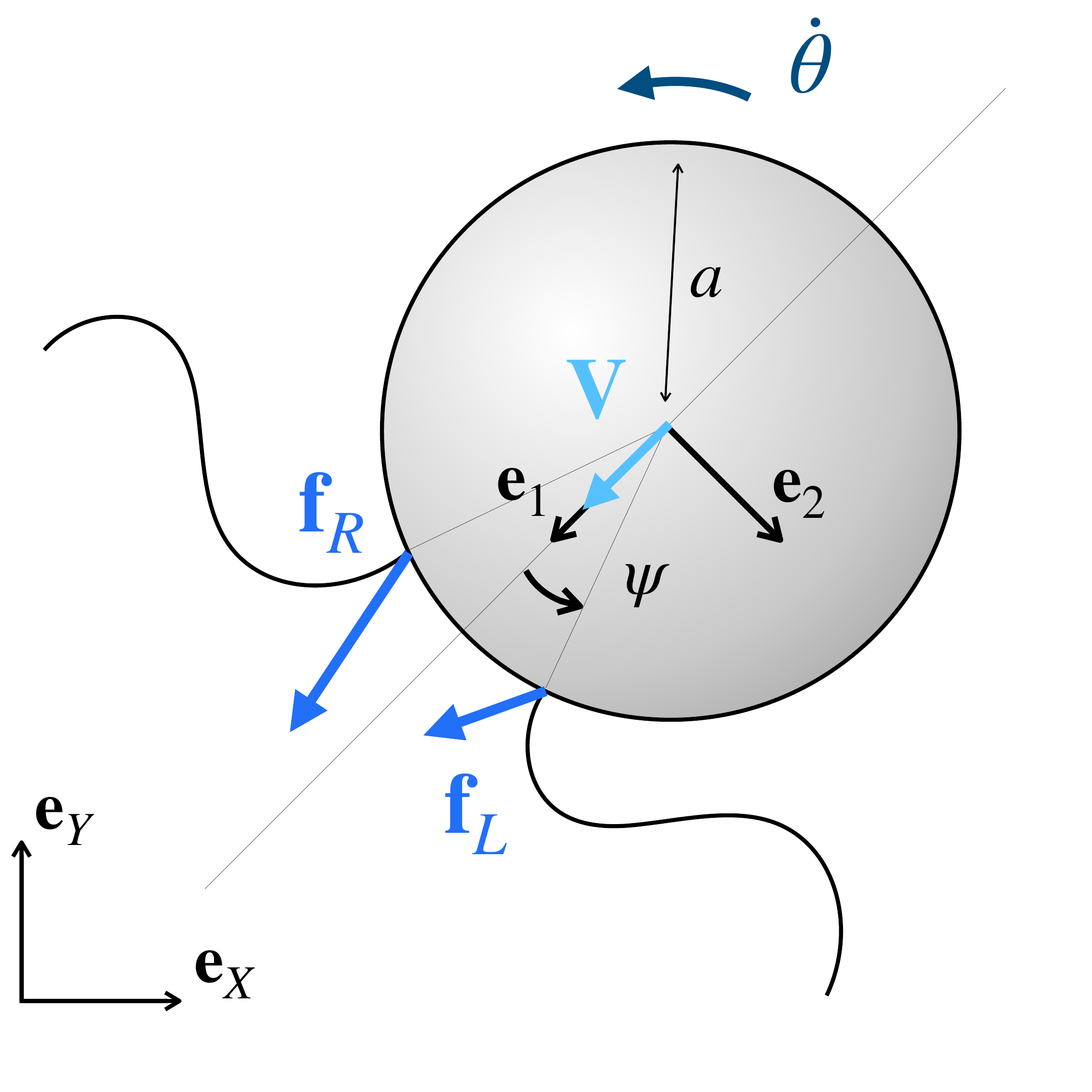}
        \put(0,100){(b)}
    \end{overpic}
    \caption{Two simple models of biflagellar propulsion, drawn to resemble an exemplar swimmer. In (a), we illustrate a minimal model that directly relates rotation rates to differences in propulsive forces, and linear velocities to sums of propulsive forces. In (b), we relax these assumptions of direct proportionality, deriving the equations of motion through a more refined force- and torque-balance argument that includes further geometrical information.}
    \label{fig: chlamy model}
\end{figure*}

In symbols, we write $(x,y)$ for the position of the swimmer in the plane and $\theta$ for its orientation relative to some fixed axis, with $x$, $y$, $\theta$ each functions of time $t$. We consider minimal models of the form
\begin{subequations}\label{eq: gen: minimal model}
\begin{align}
    \diff{x}{t} &= V\cos\theta\,,\\
    \diff{y}{t} &= V\sin\theta\,,\\
    \diff{\theta}{t} &= \eta_r\left[f_R(t) - f_L(t)\right]\,,\label{eq: gen: rotation rate}
\end{align}
\end{subequations}
where the swimming speed $V$ is given by
\begin{equation}\label{eq: gen: speed}
    V = \eta_t\left[f_R(t) + f_L(t)\right]\,.
\end{equation}
The periodic functions $f_R$ and $f_L$ represent forces generated by the right and left appendages, respectively, summarising the details of the bilateral gait. Note that their sum appears in the expression for the swimming speed, whilst their difference drives reorientation. The swimmer-dependent constants $\eta_t$ and $\eta_r$ relate the generated forces to translational and rotational velocities, respectively. Throughout, we will assume that the swimmer is progressive, so that the averages of $f_R$ and $f_L$ over one beat period are non-zero. This model is illustrated in \cref{fig: chlamy model}a.

Models of this form exhibit a range of behaviours that naturally depend strongly on the functions $f_R$ and $f_L$. In what follows, we will take $f_R$ and $f_L$ to be almost equal, differing in one of a number of ways so that we can explore the impacts of qualitatively distinct types of asymmetry. In particular, we seek to ascertain how, and to what degree, asymmetries in the ciliary propulsion impact upon the overall swimming behaviour.

\subsection*{Amplitude asymmetry}\label{sec: minimal model: amp}
We first explore the effects of asymmetry in the amplitude of the generated forces. To do so, we take $f_R=A_Rf(t)$ and $f_L=A_Lf(t)$ for positive constants $A_R$ and $A_L$, with $f(t)$ a general (periodic) function of time common to $f_R$ and $f_L$. This choice amounts to assuming that the cilia-like appendages are precisely synchronised (an assumption we later relax), with the gaits differing only in the amplitude of their propulsive effects. This leads to the system
\begin{subequations}\label{eq: amp: minimal model}
\begin{align}
    \diff{x}{t} &= V\cos\theta\,,\\
    \diff{y}{t} &= V\sin\theta\,,\\
    \diff{\theta}{t} &= \eta_r(A_R - A_L)f(t)\,,\label{eq: amp: rotation rate}
\end{align}
\end{subequations}
where the swimming speed is now
\begin{equation}\label{eq: amp: speed}
    V = \eta_t(A_R + A_L)f(t) = \frac{\eta_t(A_R+A_L)}{\eta_r(A_R-A_L)}\diff{\theta}{t}\,.
\end{equation}
With $\diff{\theta}{t}$ and $V$ differing only by a multiplicative constant, this system can be solved analytically to yield circular trajectories in the $xy$-plane. Moreover, the curvature of these trajectories can be computed as 
\begin{equation}
    \kappa = \frac{\eta_r(A_R-A_L)}{\eta_t(A_R+A_L)}
\end{equation}
whenever $A_R\neq A_L$, with the solutions becoming straight lines in the symmetric case where $A_R=A_L$.

This solution completely captures the dependence of the swimmer trajectory on the parameters and, thus, immediately generates predictions for biflagellate swimming paths. In particular, and somewhat unremarkably, asymmetry is necessary and sufficient for non-straight swimming paths. More strikingly, the curvature of these paths is entirely independent of the propulsion function $f(t)$. This provides a generalised example of the helix theorem of \citet{Shapere1989}, with amplitude asymmetry driving curved paths in a predictable and surprisingly universal way.

However, whilst in this case the analytical solution is informative, obtaining a closed-form solution will be out of reach in our later examples. Hence, in search of a more general analysis, we illustrate how one can come to the same overall conclusion without finding an exact solution. Instead, we employ a scaling argument.

\subsection*{A scaling argument} %
To make progress via a scaling argument, we focus purely on the characteristics of the long-time dynamics. In doing so, we neglect the details of the periodic propulsion function $f(t)$. With only long-term behaviour in mind, we note that \cref{eq: amp: speed} implies that the position of the swimmer evolves at a rate proportional to the sum $A_R + A_L$. Similarly, \cref{eq: amp: rotation rate} purports that the swimmer rotates at a rate proportional to the difference $A_R - A_L$. Therefore, a long-term estimate of the curvature is
\begin{equation}\label{eq: amp: scaling}
    \kappa \sim \frac{1}{V}\diff{\theta}{t} \sim \frac{A_R - A_L}{A_R + A_L}\,,
\end{equation}
up to constants that depend (in a straightforward manner) on $\eta_r$ and $\eta_t$. By construction, this curvature scaling is independent of the details of the gait encoded in $f(t)$, in line with the exact solution noted earlier. This argument effectively replaces $f(t)$ with its average over a period, so that shifts in phase can be entirely neglected, and can be made more precise with additional notation. Note that this scaling argument correctly yields the functional dependence of the curvature on the amplitudes $A_R$ and $A_L$. This includes a prediction that we omitted from our earlier discussion: whilst increasing amplitude asymmetry (i.e. increasing the difference between $A_R$ and $A_L$) leads to increasing curvatures, the growth in curvature is not unbounded. In particular, $\abs{\kappa}$ approaches a constant as $A_R/A_L\rightarrow\infty$ (and similarly for $A_L/A_R\rightarrow\infty$), where that constant depends only on the geometric details of the swimmer. In other words, the geometry of a swimmer sets a maximum possible curvature of any swimming trajectory that cannot be surpassed with any degree of amplitude asymmetry.

In summary, a simple scaling argument accurately captures key long-time dynamics of this biflagellate swimmer model, with the constructive-destructive interactions placed front and centre in the predictions of this back-of-the-envelope analysis.

\begin{figure*}
    \centering
    \begin{overpic}[width=0.49\linewidth]{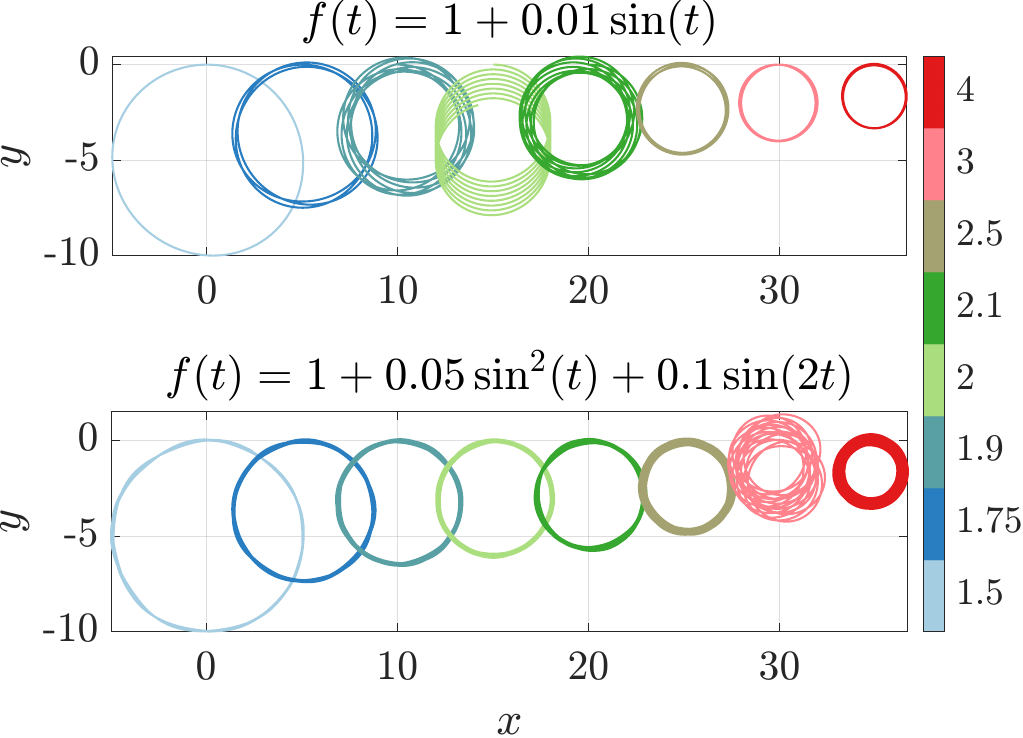}
    \put(0,70){(a)}
    \put(88,68.5){$\sfrac{k_R \!}{k_L}$}
    \end{overpic}
    \begin{overpic}[width=0.4\linewidth]{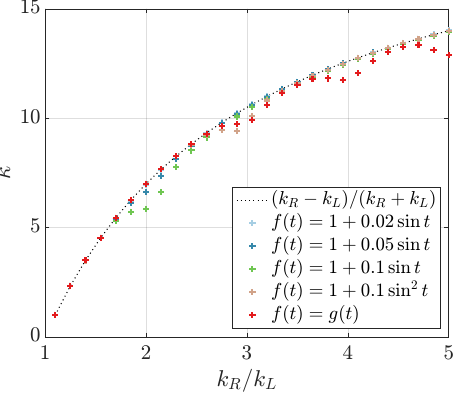}
    \put(-3,86){(b)}
    \end{overpic}
    \caption{Exploring frequency asymmetry via a scaling argument. (a) Sample trajectories in the $xy$-plane for two different functions $f$, showcasing the emergence of large-scale trajectories that are approximately circular, with resonant outliers. (b) Plots of curvature, both as computed from full numerical simulations of \cref{eq: freq: model} (coloured dots for various functions $f$) and the scaling argument of \cref{eq: freq: scaling} (dotted black line). As predicted by the scaling argument, the curvature of the swimmer paths is approximately independent of the details of the forcing function $f(t)$ -- in fact, the dots for various $f$ often coincide almost exactly for a given ratio $k_R/k_L$. Moreover, the scaling law and empirically computed curvatures show excellent agreement, validating the scaling argument.  
    In both (a) and (b), we have taken $A=1$, $\eta_r=\eta_t=1$ and $k_L = 1$, varying $k_R$. In (b), $g$ is defined as $g(t) = 1+0.05 \sin(2t)+0.05 \cos(3t)+0.05 \sin(4t)$ and we have renormalised the curvatures for each $f$ by the curvature for $k_R/k_L = 1.1$.}
    \label{fig: freq}
\end{figure*}

\subsection*{Frequency asymmetry}

Building upon these ideas, we now consider a significantly more complex source of asymmetry. Drawing inspiration from observed contrasts between the cis and trans cilia of \textit{Chlamydomonas} when turning, we investigate the effects of frequency asymmetries \cite{Wan2014}. To do so, we consider \cref{eq: gen: minimal model} and take $f_R(t) = Ak_Rf(k_R t)$ and $f_L(t) = Ak_L f(k_L t)$. Here, $k_R$ and $k_L$ are constants, $A$ is a shared (constant) amplitude parameter, and $f$ is again an unspecified periodic function. Note that the frequency multipliers $k_R$ and $k_L$ also contribute to the magnitude of the propulsion in this case, as follows from the linearity and time invariance of inertia-free swimming \cite{lauga2020fluid}. Altogether, this leads to the modified swimming model
\begin{subequations}
\begin{align} \label{eq: freq: model}
    \diff{x}{t} &= V\cos\theta\,,\\ 
    \diff{y}{t} &= V\sin\theta\,,\\
    \diff{\theta}{t} &= A\eta_r\left[ k_Rf(k_R t) - k_Lf(k_L t)\right]\,, \label{eq: freq: modelc}
\end{align}
\end{subequations}
where the swimming speed is given by
\begin{equation}
    V = A\eta_t\left[k_Rf(k_R t) + k_Lf(k_L t)\right].
    \label{eq: freq: speed}
\end{equation}
Notably, the swimming speed is no longer proportional to the rate of rotation. In \cref{app: eff resistance argument}, we discuss how an effective model of this form can be derived from more detailed models.

In this case, an exact solution is less forthcoming. Instead, we apply a scaling argument analogous to that described above to obtain the curvature relation
\begin{equation}\label{eq: freq: scaling}
    \kappa \sim \frac{k_R - k_L}{k_R + k_L}\,,
\end{equation}
once again effectively averaging over gait fluctuations and omitting swimmer-dependent constants. Not only do we recover a maximum capacity for curved swimming that is set by the geometrical details of the swimmer in the limit of large $k_R$ or $k_L$, we also predict the functional dependence on $k_R$ and $k_L$. Given our assumptions, this result is universal, in the sense that curvatures arising from frequency differences in biflagellate swimming are predicted to follow the fractional linear relationship of \cref{eq: freq: scaling}, independent of the details of the swimmer and its gait.

Before considering this model through the lens of a more refined analysis to further motivate and justify the approach taken here, we examine our prediction numerically. \Cref{fig: freq} reports the results of a wide variety of numerical realisations of the model of \crefrange{eq: freq: model}{eq: freq: speed}, varying both $f$ and the ratio $k_R/k_L$. In \cref{fig: freq}a, we superimpose sample trajectories for two choices of oscillatory $f$, highlighting a marked tendency for circular swimming paths that is in line with the prediction of constant curvature. With this partial validation, we repeat analogous numerical explorations for a broader range of $f$ and $k$ and estimate the curvature of the paths by fitting the trajectories to circles. These curvatures are plotted alongside the prediction of \cref{eq: freq: scaling} in \cref{fig: freq}b, where we have normalised the former by the curvature when $k_R/k_L = 1.1$ and fixed $A=\eta_r=\eta_t=k_L=1$. Remarkably, with minor exceptions that we explore further below, the numerical results collapse onto the analytical prediction, showing little to no deviation from the expected relationship. This observation appears to apply across the considered $f$ and a wide range of $k_R/k_L$ and serves as excellent validation of the scaling argument.

Though broadly validating the predictions of the scaling argument, \cref{fig: freq} does reveal occasional departures from the expected behaviour, seemingly localised around integer values of $k_R/k_L$. Close to these values, both the circularity of the trajectories and the precise agreement between the numerically computed curvature and the prediction of \cref{eq: freq: scaling} are, to some extent, lost, though not universally. The fluctuations appear to be somehow related to the base frequency of the gait function $f(t)$: we note how taking $f(t)=g(t)$, as defined in the caption of \cref{fig: freq} and composed of many distinct oscillatory modes, gives rise to prominent fluctuations for larger integer values of $k_R/k_L$, in contrast to those observed for the other considered $f(t)$. In order to explore this in more detail, along with placing the rest of our simplified analysis on a firmer theoretical footing, we next consider a more complex model of biflagellate swimming.

\section{Improved swimmer model}\label{sec: better model}
In posing and exploring our minimal model above, we have knowingly neglected details of the swimmer and the swimming mechanism that are potentially significant. Here, we refine our approach by considering a more detailed model that incorporates some degree of swimmer geometry and a more general description of biflagellar forcing. 

\subsection*{Formulation}\label{sec: better model: formulation}
As before, we consider a swimmer moving in the $xy$-plane and denote the location of the swimmer centre by coordinates $(x,y)$, which we now write in vector form as $\vec{X} = [x,y]^T$. Here, we explicitly define the swimmer orientation $\theta$ as the angle between a fixed laboratory frame orthonormal basis $\{\vec{e}_X,\vec{e}_Y\}$ and a swimmer-fixed orthonormal basis $\{\vec{e}_1,\vec{e}_2\}$, with $\vec{e}_1 = \cos{\theta}\vec{e}_X + \sin{\theta}\vec{e}_Y$. Two motile cilia-like appendages are positioned symmetrically about $\vec{e}_1$ at angle $\psi$, and are entirely contained within the plane. This setup and notation is illustrated in \cref{fig: chlamy model}b.

Analogously to \cref{sec: minimal model}, the forces exerted on the body by each cilium are denoted by $\vec{f}_R$ and $\vec{f}_L$. We then decompose these forces into components $f_R^{(1)}$, $f_R^{(2)}$, $f_L^{(1)}$, and $f_L^{(2)}$ along the basis directions of the swimmer-fixed frame, with
\begin{equation}\label{eq: coordinates force first cilium}
    \vec{f}_R = f_R^{(1)} \vec{e}_1 + f_R^{(2)} \vec{e}_2
\end{equation}
    and
\begin{equation}\label{eq: coordinates force second cilium}
    \vec{f}_L = f_L^{(1)} \vec{e}_1 - f_L^{(2)} \vec{e}_2\,.
\end{equation}
We highlight our convention that the force contributions in the $\vec{e}_1$ direction have the same sign, whilst those along $\vec{e}_2$ appear with opposite signs. This convenient choice entails that we could model a mirror-symmetric biflagellate by taking $f_R^{(1)}=f_L^{(1)}$ and $f_R^{(2)}=f_L^{(2)}$, with $\vec{e}_1$ then being the emergent swimming direction.

With this notation, the equations of force and moment balance on the swimmer become
\begin{subequations}\label{eq: better model}
\begin{align}
    6\pi\mu a \dot{\vec{X}} &= [f_R^{(1)} + f_L^{(1)}]\vec{e}_1 + [f_R^{(2)} - f_L^{(2)}]\vec{e}_2 \,,\\
    8\pi\mu a^3\dot{\theta} &= [f_R^{(1)} - f_L^{(1)}]\sin{\psi} +[f_R^{(2)} - f_L^{(2)}]\cos{\psi}\,,
\end{align}
\end{subequations}
having applied Stokes' law on the sphere with radius $a$ for a fluid of viscosity $\mu$. Note that, as $\vec{e}_1=\cos{\theta}\vec{e}_X + \sin{\theta}\vec{e}_Y$, this model collapses onto that of \cref{sec: minimal model} if we neglect the $\vec{e}_2$ components of the ciliary forcing.

Before we proceed to analyse this model, it is worth acknowledging a key implicit assumption. In writing \cref{eq: better model}, we are assuming that the forcing terms are independent of the linear and angular speed of the swimmer; in actuality, forces are generated by ciliary motion relative to the motion of the swimmer. We present an estimate of the magnitude of the associated errors in \cref{app: errors in better model}, and will later validate the predictions of this model via a more detailed computational approach.

\subsection*{Validating the minimal model}
With the model of \cref{eq: better model}, we examine the validity of the scaling law identified in \cref{eq: freq: scaling} for the curvature of the swimming path as a function of the frequency difference. A similar calculation for the case of amplitude asymmetry is presented in \cref{app: better model: amplitude}.

First, let us decompose $f_R^{(1)}$ and $f_R^{(2)}$ into average and oscillatory components, writing
\begin{equation}\label{eq: better model: f_R}
    f_R^{(i)} = k_R[A^{(i)} + B^{(i)}J(k_R t)]
\end{equation}
for constants $A^{(i)}$, $B^{(i)}$, and $k_R$, periodic function of time $J(k_R t)$ with zero mean, and $i\in\{1,2\}$. We capture frequency asymmetry via a similar decomposition of $f_L^{(1)}$ and $f_L^{(2)}$, with
\begin{equation}\label{eq: better model: f_L frequency asymmetry}
    f_L^{(i)} = k_L[A^{(i)} + B^{(i)}J(k_L t)]
\end{equation}
having substituted $k_L$ for $k_R$ and kept all else equal. Defining constants
\begin{equation}
    \alpha \coloneqq \frac{A^{(1)}\sin{\psi} + A^{(2)}\cos{\psi}}{8\pi\mu a^2}\,, \quad 
    \beta \coloneqq \frac{B^{(1)}\sin{\psi} + B^{(2)}\cos{\psi}}{8\pi\mu a^2}
\end{equation}
which represent aggregated magnitudes of the average and oscillatory components of the ciliary forcing, respectively, we can write the angular evolution of \cref{eq: better model}b as
\begin{equation}
    \dot{\theta} = \alpha[k_R-k_L] + \beta [k_R J(k_R t) - k_L J(k_L t)]\,.
\end{equation}
Integrating immediately gives
\begin{equation}\label{eq: better model: theta exact}
    \theta = \alpha[k_R - k_L]t + \beta[K(k_R t) - K(k_L t)]\,,
\end{equation}
where $K$ is an antiderivative of $J$ and we have assumed that $\theta(0)=0$ without loss of generality. Note that $K$ is bounded as $J$ is periodic with zero mean.

If we assume that the average effect of the ciliary propulsion is greater than the magnitude of oscillations around that average, so that $\abs{B^{(1)}},\abs{B^{(2)}}\ll\abs{A^{(1)}},\abs{A^{(2)}}$ and, thus, $\abs{\beta} \ll \abs{\alpha}$, we arrive at the approximate solution
\begin{equation}\label{eq: better model: approximate theta}
    \theta(t)  = \alpha [k_R - k_L]t\,, 
\end{equation}
corresponding to rotation over a period of $2\pi/\alpha [k_R - k_L]$.

We can make similar progress with the linear velocity equation. Dropping terms involving $B^{(1)}$ and $B^{(2)}$ in favour of terms scaling with $A^{(1)}$ and $A^{(2)}$, we arrive at the approximate system
\begin{equation}
    6\pi\mu a \begin{pmatrix}
        \dot{x}\\ \dot{y}
    \end{pmatrix}
    =
    \begin{pmatrix}
        \cos{\theta} & -\sin{\theta}\\
        \sin{\theta} & \cos{\theta}
    \end{pmatrix}
    \begin{pmatrix}
        [k_R+k_L]A^{(1)}\\ [k_R-k_L]A^{(2)}
    \end{pmatrix}\,.
\end{equation}
With $\theta$ approximated by \cref{eq: better model: approximate theta} and assuming $k_R \neq k_L$, we can integrate this directly to give
\begin{equation}
    6\pi\mu a \alpha[k_R-k_L] \begin{pmatrix}
        x\\ y
    \end{pmatrix}
    =
    \begin{pmatrix}
        \sin{\theta} & \cos{\theta}\\
        -\cos{\theta} & \sin{\theta}
    \end{pmatrix}
    \begin{pmatrix}
        [k_R+k_L]A^{(1)}\\ [k_R-k_L]A^{(2)}
    \end{pmatrix}\,,
\end{equation}
setting constants of integration to zero without loss of generality. This corresponds to circular motion with (unsigned) curvature
\begin{equation}\label{eq: better model: full curvature relation}
    \kappa = \frac{6\pi\mu a\abs{\alpha}\abs{k_R-k_L}}{\sqrt{\left([k_R+k_L]A^{(1)}\right)^2 + \left([k_R-k_L]A^{(2)}\right)^2}}
\end{equation}
Hence, under the relatively weak assumption that the average propulsion dominates oscillations around that average, we see that this more complex model reproduces constant-curvature motion.

Moreover, if we suppose that, on average, propulsion is predominantly along the axis of symmetry of the swimmer, we can further reduce \cref{eq: better model: full curvature relation}. In symbols, assuming that $\abs{[k_R-k_L]A^{(2)}}\ll\abs{[k_R+k_L]A^{(1)}}$, which is expected to hold in all but the most extreme cases, 
 this gives
\begin{equation}
    \kappa = \frac{3}{4a}\abs{\sin{\psi} + \frac{A^{(2)}}{A^{(1)}}\cos{\psi}} \abs{\frac{k_R-k_L}{k_R+k_L}}\,.
\end{equation}
This leading-order curvature relation agrees with the prediction of the minimal model of \cref{eq: freq: scaling}, thus serving as excellent validation of the back-of-the-envelope approach of \cref{sec: minimal model}. An entirely analogous calculation reveals the same agreement for the case of amplitude asymmetry.

In summary, under the assumption that the average propulsion dominates oscillations around that average, this more complex model reproduces circular motion. Furthermore, given the additional assumption of principally forwards propulsion (on average), we recover the form of the scaling law predicted by the minimal model.

\subsection*{Resonant frequency asymmetries}
In \cref{sec: minimal model}, we noted the appearance of small but non-trivial differences between the predictions of the scaling argument and simulations around integer values of $k_R/k_L$. With our improved swimmer model, we are in a position to explain these discrepancies.

In the validation above, we assumed that $\abs{\beta}\ll\abs{\alpha}$ and, hence, neglected the terms involving $\beta$ in \cref{eq: better model: theta exact}.  This allowed us to straightforwardly integrate the translation equations to give circular trajectories, incurring errors that scale with $\beta$. However, for $\beta$ small but non-zero, the contribution of the term $\beta[K(k_R t) - K(k_L t)]$ to the translational motion over time need not be small. To see this, it is important to highlight that, due to the nonlinearity of $\cos\theta$ and $\sin\theta$, small \emph{relative} errors in $\theta$ need not lead to small errors in $\cos{\theta}$, $\sin{\theta}$, and, crucially, their integrals.

\begin{figure}
    \centering
    \includegraphics[width=0.6\linewidth]{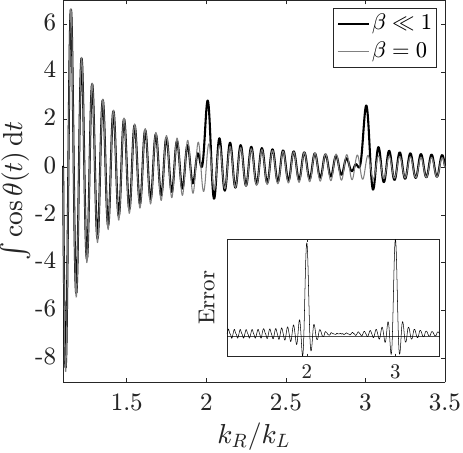}
    \caption{Resonance accompanies frequency asymmetries, explaining discrepancies in the predicted scaling law. We plot the integral of $\cos{\theta(t)}$, which contributes to one component of the swimmer translation, for $\beta\ll1$ (black) and $\beta=0$ (grey). The approximation of \cref{eq: better model: approximate theta} corresponds to $\beta=0$, which closely matches the exact result with $\beta\ll1$ except near $k_R/k_L\in\{2,3\}$, where resonance occurs and the approximation incurs significant errors. Inset, we show the difference between the two curves, which is small apart from near the resonant frequencies. Here, we have taken $J(z) = \cos(z) + \cos(2z)$ and $\beta/\alpha=0.05$ and integrated from $t=0$ to $t=30\pi$.}
    \label{fig: resonant freq asymmetry}
\end{figure}

\begin{figure*}
    \centering
    \begin{overpic}[width=0.32\linewidth]{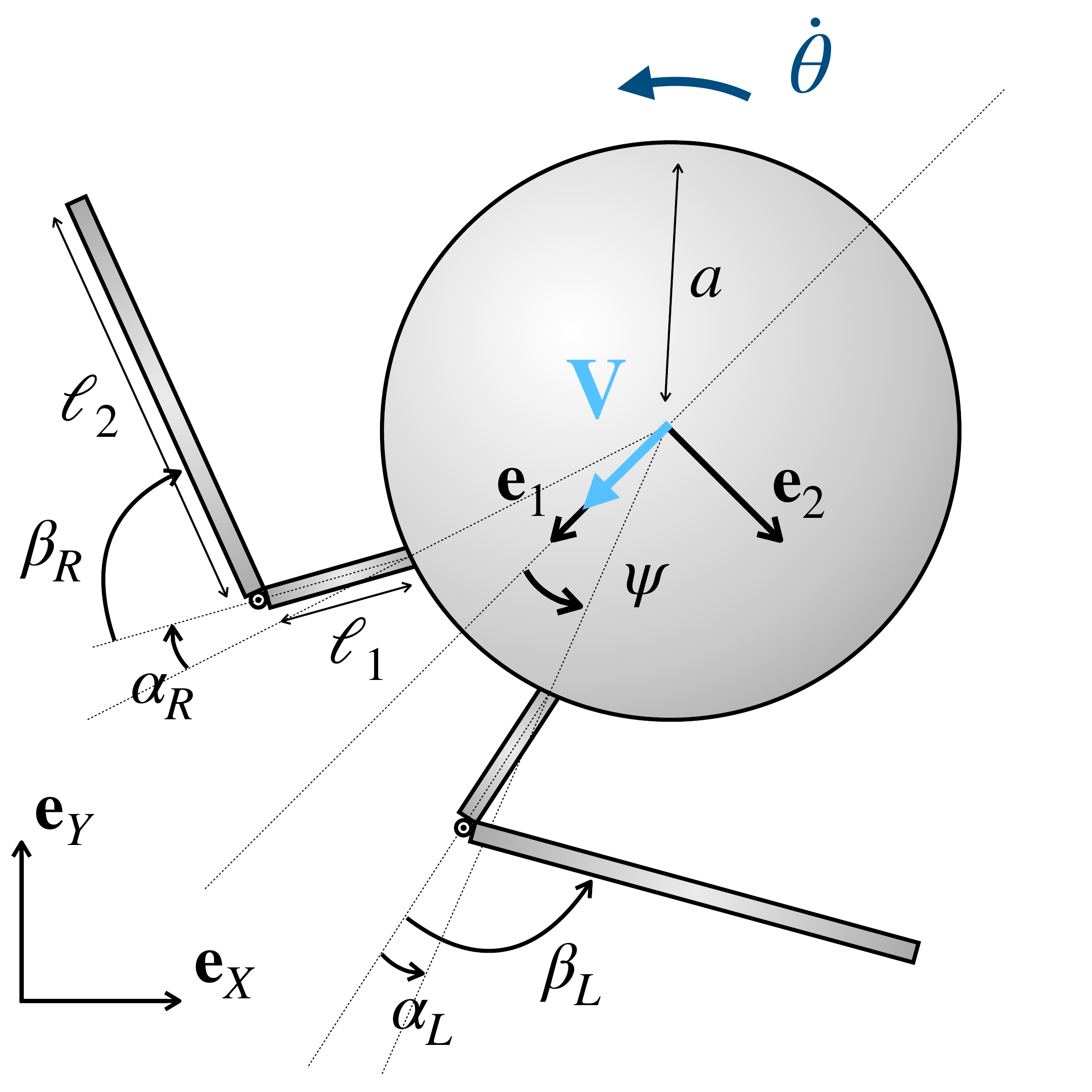}
        \put(0,107){(a)}
    \end{overpic}
    \hspace{2em}
    \begin{overpic}[width=0.4\linewidth]{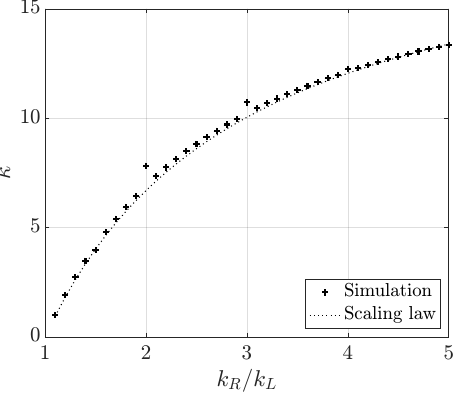}
        \put(-3,86){(b)}
    \end{overpic}
    \caption{Validating the frequency-asymmetry scaling law with a detailed computational model. Using the detailed model of \cref{sec: computational model}, illustrated in (a), we repeat the exploration of \cref{fig: freq} for a prescribed beating pattern. (b) Varying $k_R/k_L$, we simulate the motion and compute the resulting path curvatures (black crosses). We also plot the prediction of the scaling law of \cref{eq: freq: scaling} with a fitted constant of proportionality (black dotted curve), evidencing excellent agreement with the computational results. As predicted by the analysis of our improved swimmer model, resonance-like deviations from the prediction can be seen for discrete values of $k_R/k_L$.}
    \label{fig: computational model}
\end{figure*}

To see this concretely, consider $J(z) = \cos(z) + \cos(2z)$ and  $\beta/\alpha = 0.05$. Fixing $k_R$  and varying the ratio $k_R/k_L$, we plot the numerically estimated integral of $\cos{\theta}$ over $t\in\{0,30\pi\}$ in \cref{fig: resonant freq asymmetry} as a black curve, with $\alpha$ also fixed and $\theta$ given by \cref{eq: better model: theta exact}.  We overlay the analogous integral when $\beta=0$ as a grey curve, which corresponds to the approximation of \cref{eq: better model: approximate theta}. These curves are in excellent agreement except when $k_R/k_L$ is close to 2 or 3; notably, these are precisely the locations of the discrepancies observed in \cref{fig: freq}b. For frequency ratios near to these values, the contributions of the $\beta$ terms to the translation cannot be ignored; doing so leads to errors that accumulate over time. More generally, the same phenomenon occurs whenever the frequency difference $\abs{k_R-k_L}$ is approximately equal to any frequency present in a Fourier decomposition of $K(k_R t)$ or $K(k_L t)$. If $c$ is the coefficient of such a term, errors in the predicted translation will scale with $c\beta T/2$ over a timescale $T$. This can be seen explicitly by expanding out $\cos{\theta}$ and $\sin{\theta}$ to isolate the terms involving $\beta$.

Hence, overall, we have seen that the more complex model introduced in this section broadly validates the conclusions of the simpler model of \cref{sec: minimal model} and explains the origins of the noted discrepancies between the scaling laws and numerical simulations in specific cases.

\section{Detailed computational model}\label{sec: computational model}
Both of the models considered thus far have neglected the deformation of the swimmer, focusing on the effective force generated on the body by flagellar deformation. In this section, we verify the predictions of these simplified models using a detailed computational model that explicitly accounts for the motion of two cilia-like appendages.

\subsection*{Formulation}
We model the swimmer as a rigid sphere, connected to which are two actively moving model cilia, as shown in \cref{fig: computational model}a. Each cilium is composed of two straight line segments of fixed length connected by a hinge, with the cilium itself hinged at the base where it attaches to the spherical body. To model motility, we prescribe the evolution of the angles $\alpha_i$ and $\beta_i$ subtended at these hinges, with $i\in\{R,L\}$ indexing the two cilia and choosing the time dependence such that they exhibit features characteristic of biological examples. We utilise the resistive force theory of \citet{Gray1955,Hancock1953} to relate the velocity of the cilia to the forces exerted on them by the ambient fluid medium, exploiting the slenderness of the cilia (here with aspect ratio 1:1000) and incurring logarithmic errors in general. Solving the governing equations of force and torque balance yields the motion of the swimmer, which we constrain to be in the plane by prescribing planar deformation of the cilia and attaching them at the equator of the spherical body. We omit the details of the governing equations for brevity; computer codes for simulating this swimmer model are provided \cite{githubRepo} and contain all requisite information.

\begin{figure*}[t]
    \centering
    \begin{overpic}[width=0.3\linewidth]{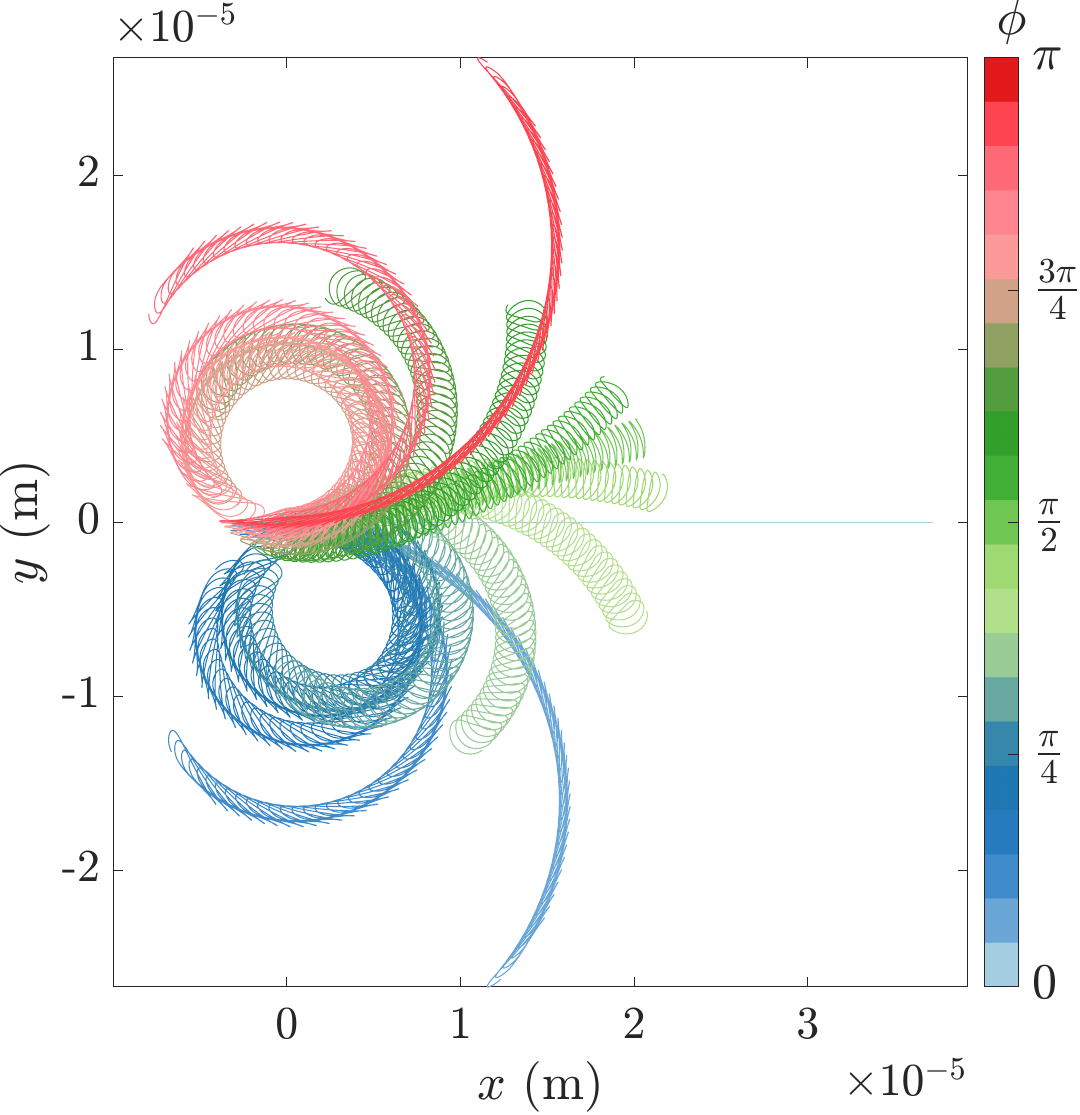}
    \put(0,100){(a)}
    \end{overpic}
    \begin{overpic}[width=0.3\linewidth]{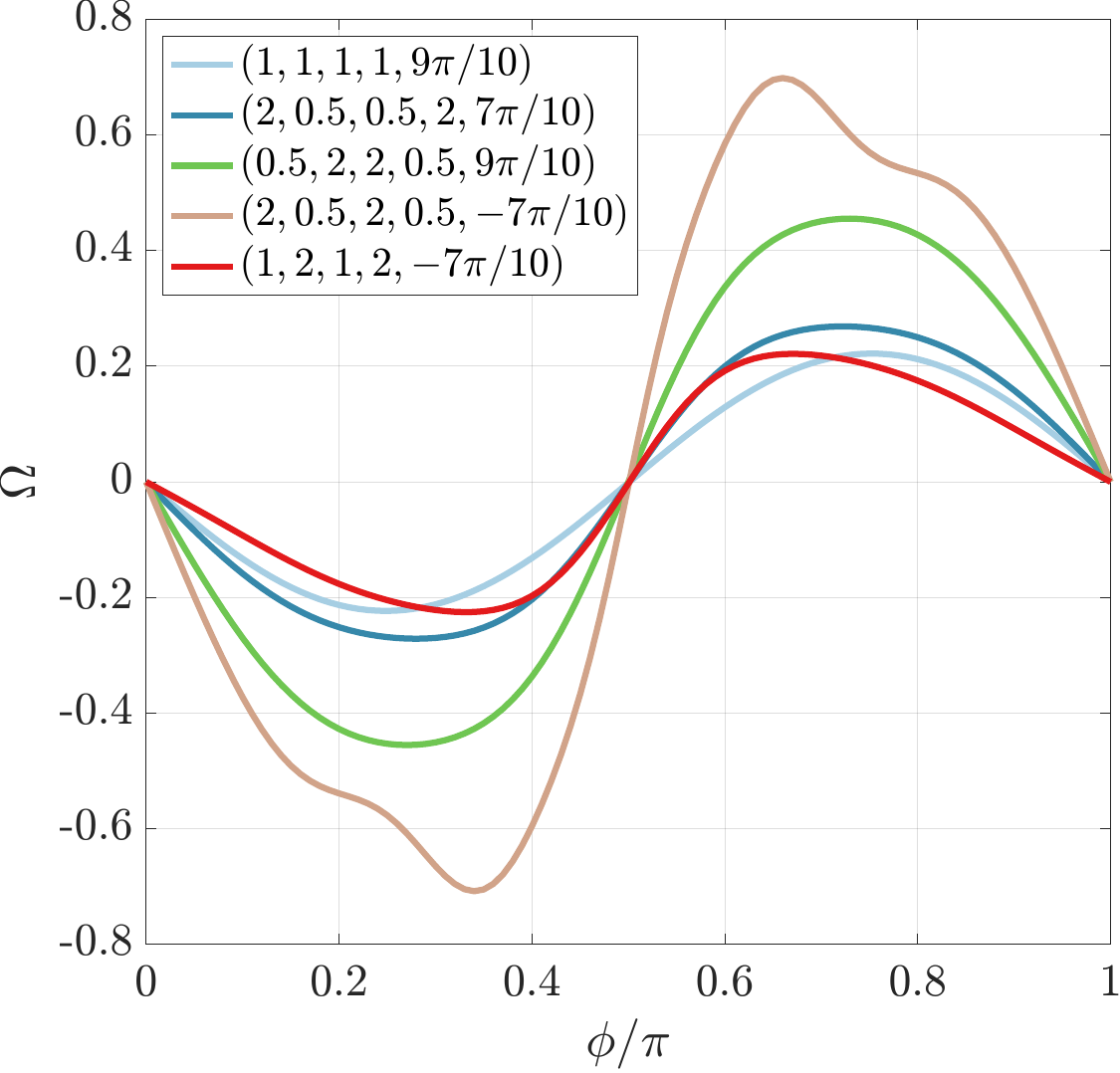}
    \put(0,100){(b)}
    \end{overpic}
    \begin{overpic}[width=0.29\linewidth]{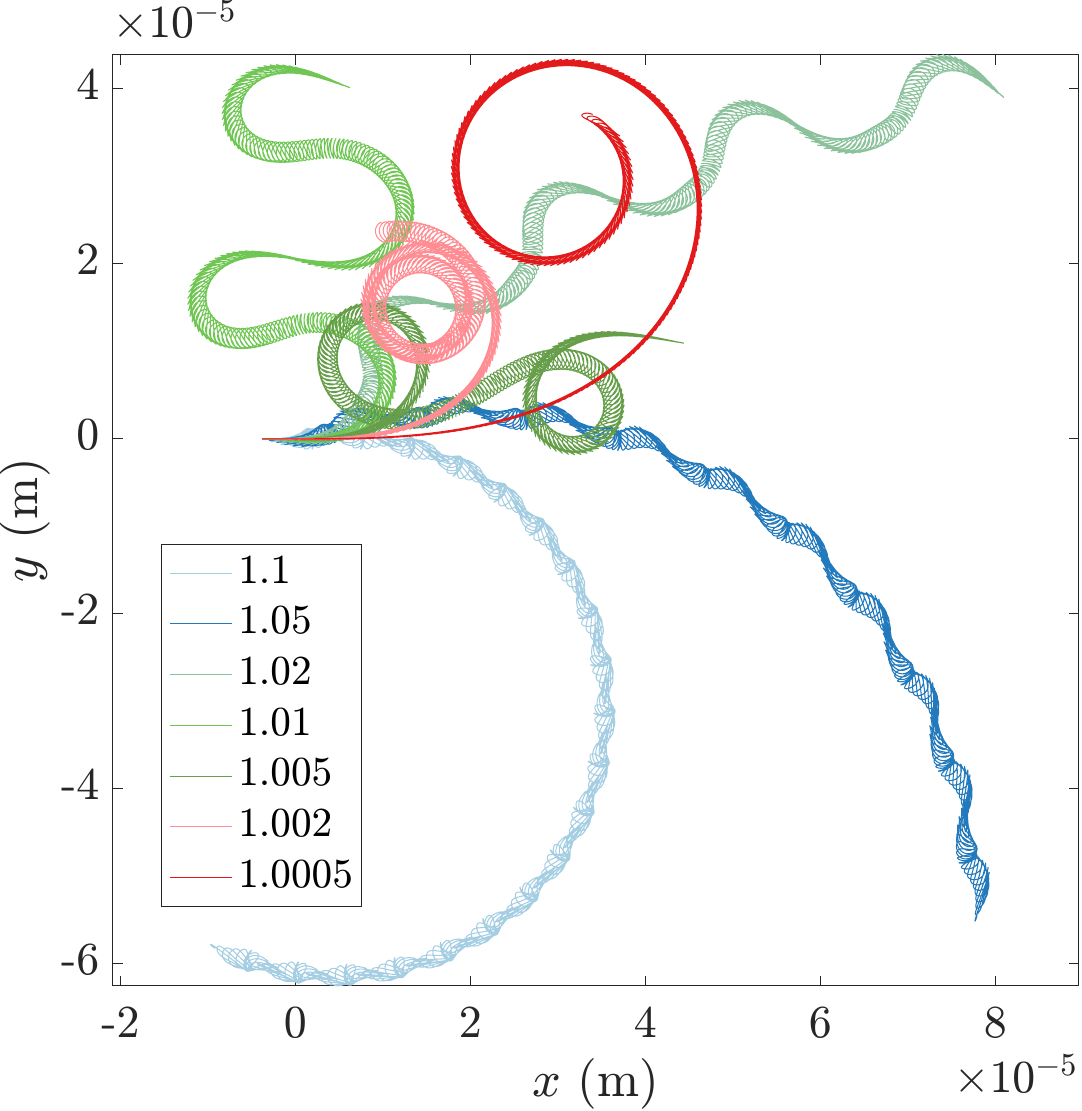}
    \put(0,100){(c)}
    \end{overpic}
    \caption{Exploring phase asymmetry. (a) Emergent trajectories for a range of phase shifts $\phaseshift$ (indicated with colour), with neither amplitude nor frequency asymmetry. Phase differences can be seen to readily generate curved paths, except for $\phi = 0$ and $\phi = \pi$, which yield straight lines overlaid on top of each other. (b) Average angular velocity $\Omega$ as a function of phase shift $\phaseshift$, with different curves representing different gaits (specifically, we set $\alpha(t) = a_0  + a_1 \sin(t)$, and $\beta(t) = b_0 + b_1 \cos(t + \varphi)$. The legend reports each parameter set as $(a_0,a_1,b_0,b_1,\varphi)$). This highlights a complex interplay between gait and phase shift, with gait-dependent differences in even the $\phaseshift$ at which maximal swimmer rotation occurs. (c) Trajectories on long timescales for very small frequency differences show failure of the frequency scaling when the frequency ratio is smaller than 1.05, with the emergence of complex behaviours due to interplay between phase shift and frequency difference.  }
    \label{fig: phase}
\end{figure*}

\subsection*{Validating simpler models}
Using this more detailed computational model, we repeat the explorations of the previous sections to validate our conclusions in a system that more faithfully represents a biflagellated swimmer. We prescribe $\alpha_i=\alpha(k_i t)$ and $\beta_i=\beta(k_i t)$ for functions $\alpha$ and $\beta$ (given in \cref{app: alpha beta}) such that the resulting beating patterns are reminiscent of the ciliary beating of \textit{Chlamydomonas reinhardtii}, focusing on exploring differences in frequency between the two cilia. %

Repeating the investigation into the role of frequency asymmetry in driving curved swimming, in \cref{fig: computational model}b we plot the curvature of simulated swimming paths using this computational model alongside the fitted scaling law of \cref{eq: freq: scaling}. We observe remarkable agreement between the scaling law and the computed curvatures, further validating the predictions of the previous models. Furthermore, we also recover the resonance-like phenomenon observed in the previous section, with clear deviations from the prediction of the simpler models arising for $k_R/k_L\in\{2,3,4\}$ in this case. Thus, the simple models analysed here successfully capture the essential mechanisms driving the emergence of curved swimming paths in this detailed model of biflagellate locomotion.

\subsection*{Exploring phase asymmetry}
This most detailed model also allows us to explore another mechanism for symmetry breaking: phase asymmetry. Explicitly, we consider the effects of taking $\alpha_i=\alpha(kt + \phaseshift_i)$ and $\beta_i=\beta(kt + \phaseshift_i)$, where $\phaseshift_R=0$ and $\phaseshift_L=\phaseshift$ is a constant phase shift. This qualitatively resembles reported phase differences between the \textit{cis} and \textit{trans} cilia of \textit{C. reinhardtii} \cite{wang2025light}.

Simulating this phase-shifted swimmer for the same $\alpha$ and $\beta$ as above yields the trajectories shown in \cref{fig: phase}a. As might be expected, phase asymmetry gives rise to curved trajectories, with the curvature dependent on the phase difference. However, in contrast to our explorations of amplitude and frequency asymmetry, the dependence of the curvature of these paths on the phase difference appears far from simple, even depending sensitively on the details of the gait. To illustrate this, we plot the emergent angular velocity as a function of phase shift $\phaseshift$ in \cref{fig: phase}b, demonstrating a complex dependence of the dynamics on $\phaseshift$ that varies between gaits (different curves correspond to different gaits).

The complexity of this dependence is reflected by the fact that the simpler models are not able to capture it in their presented forms. In the minimal model of \cref{eq: gen: minimal model}, a phase asymmetry corresponds to taking $f_R(t)=f(t)$ and $f_L(t)=f(t+\phaseshift)$ for some periodic function $f$ and constant phase shift $\phaseshift$. Averaging \cref{eq: gen: rotation rate} over a period yields zero, so that the net rotation of the swimmer as predicted by this model is zero. Analogous reasoning leads to the same conclusion in the improved model of \cref{eq: better model}.

The key factor missing from the simpler models is the time evolution of the rotational resistance of the swimmer, with us having taken $\eta_r$ to be constant in the previous sections. In the computational model, this resistance depends explicitly on the instantaneous swimmer shape and, therefore, varies throughout the motion. By introducing time and phase-shift dependence to $\eta_r$ in the simpler models, the emergence of curved trajectories can be predicted, though no simple scaling laws are apparent. We perform an explicit calculation in \cref{app: phase shift} that highlights how a phase shift gives rise to net rotation in a special case of the computational model, where state-dependent resistance is accounted for.

Remarkably, this configuration-dependent resistance effect has little impact on the cases of frequency and amplitude asymmetry; indeed, we neglected changes in resistance in our earlier models and nevertheless observed excellent agreement with the computational model (\cref{fig: computational model}b). Focusing on the case of frequency asymmetry, this can explained by considering the periods of the forcing terms $f(k_R t)$, $f(k_L t)$ and the instantaneous resistance coefficients $\eta_r(t,k_R,k_L)$, $\eta_t(t,k_R,k_L)$, with the latter depending on the swimmer configuration and, thus, both $k_R$ and $k_L$. We present a detailed argument in \cref{app: eff resistance argument}, in which we derive effective resistance constants from the full computational model.

If we focus on only slight frequency asymmetries, the relative configurations of the two cilia and, therefore, the resistance to rotation and translation evolve on a slow timescale of $\abs{k_R-k_L}^{-1}$. In this case, the dynamics quasi-statically resemble those corresponding to a phase shift, with that shift evolving slowly over many cycles. This drifting phase asymmetry is expected to lead to dynamics that are more complex than we have observed thus far. A numerical exploration that varies the ratio $k_R/k_L$ is presented in \cref{fig: phase}c, demonstrating the wide range of behaviours that can result from very slight frequency asymmetries.

\section{Robophysical realisation}
Our theoretical models predict a simple generalised scaling law relating trajectory curvature to asymmetry in a biflagellate microswimmer. 
To validate this functional relationship in a physical context, we conceived of and implemented a macroscale, biflagellated, fully autonomous robophysical model (based on a previous design \cite{diaz2021minimal,moreau2024minimal}) capable of locomotion and swimming at low Reynolds number. 
Individual robotic `cilia' are constructed from a series of 3D printed segments connected by hinges, to generate an asymmetric stroke cycle. 
Each cilium is actuated by a waterproof servo motor (IP68 digital micro-servo SW-0250 MG, SAVOX) and attached to a cylindrical PLA hull (approximately \SI{8}{\centi\metre} diameter, \SI{6}{\centi\metre} tall), which constitutes the body of the robotic flagellate. 
A \SI{5}{\volt} / \SI{16}{\mega\hertz} Adafruit Pro Trinket generates independent 16-bit PWM signals for the servos. Power is provided by a two-cell Li-Po pack regulated through a \SI{3}{\ampere} buck converter to maintain a constant \SI{6}{\volt} on the servo rail, while the controller draws directly from the raw battery.
All electronics are fully enclosed in a sealed cabin.

\begin{figure*}
    \centering
    \begin{overpic}[width=0.9\textwidth]{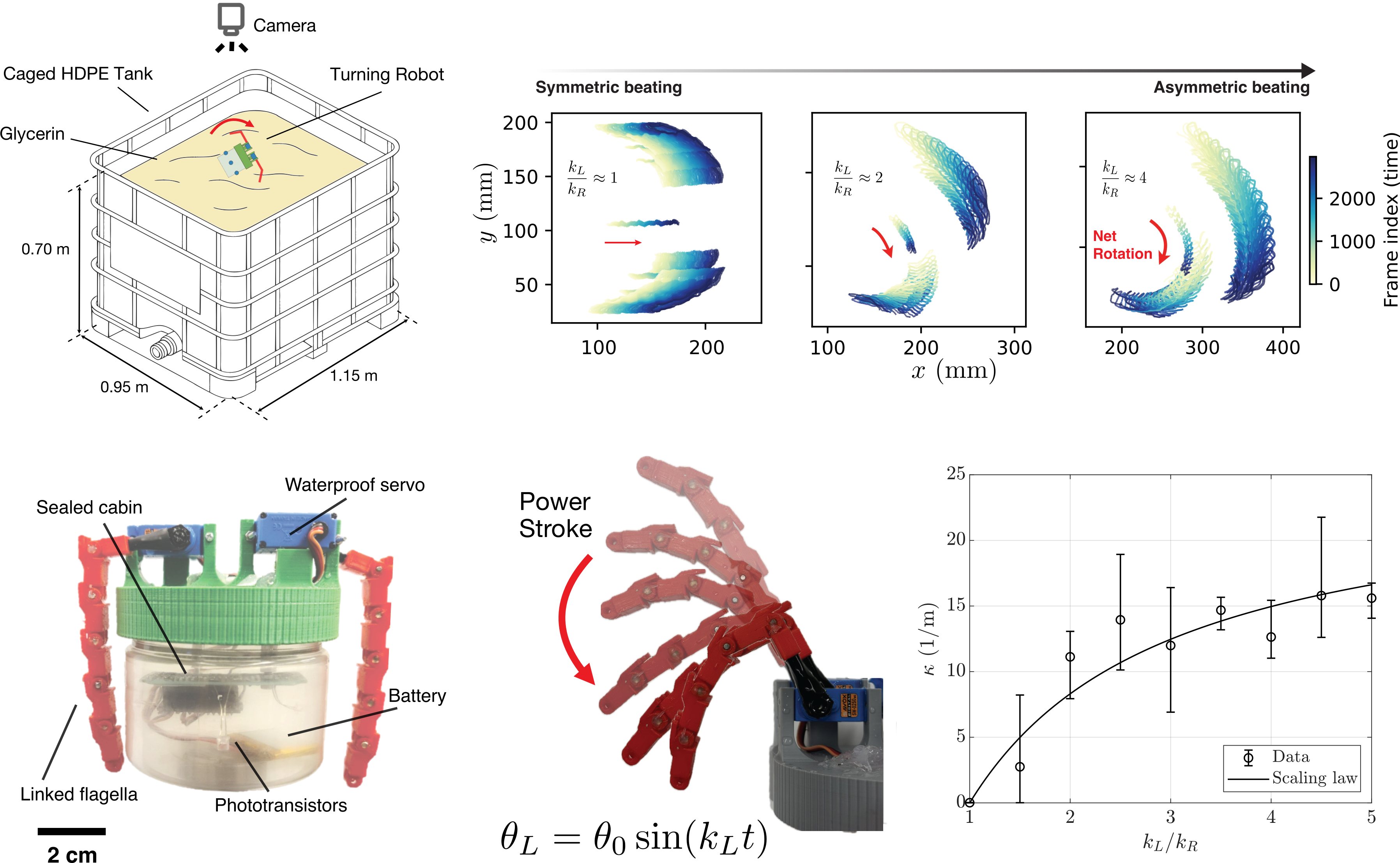}
        \put(0,60){(a)}
        \put(35,60){(b)}
        \put(0,30){(c)}
        \put(35,30){(d)}
        \put(65,30){(e)}
    \end{overpic}
    \caption{Robophysical model to probe frequency asymmetry. (a) Schematic of the experimental system. (b) Tracked positions of reference points on the robotic flagella, showing that frequency asymmetry leads to increasingly curved tracks. (c) Close-up view of the macroscale robophysical model, with two flagella-like appendages consisting of 3D-printed links. (d) Illustration of the beat generated by each appendage, actuated via the base at a prescribed frequency. (e) Fit of the scaling law to experimentally measured curvatures, showing the mean (circles) and range (bars) and demonstrating good agreement within experimental error.}
    \label{fig: robot}
\end{figure*}
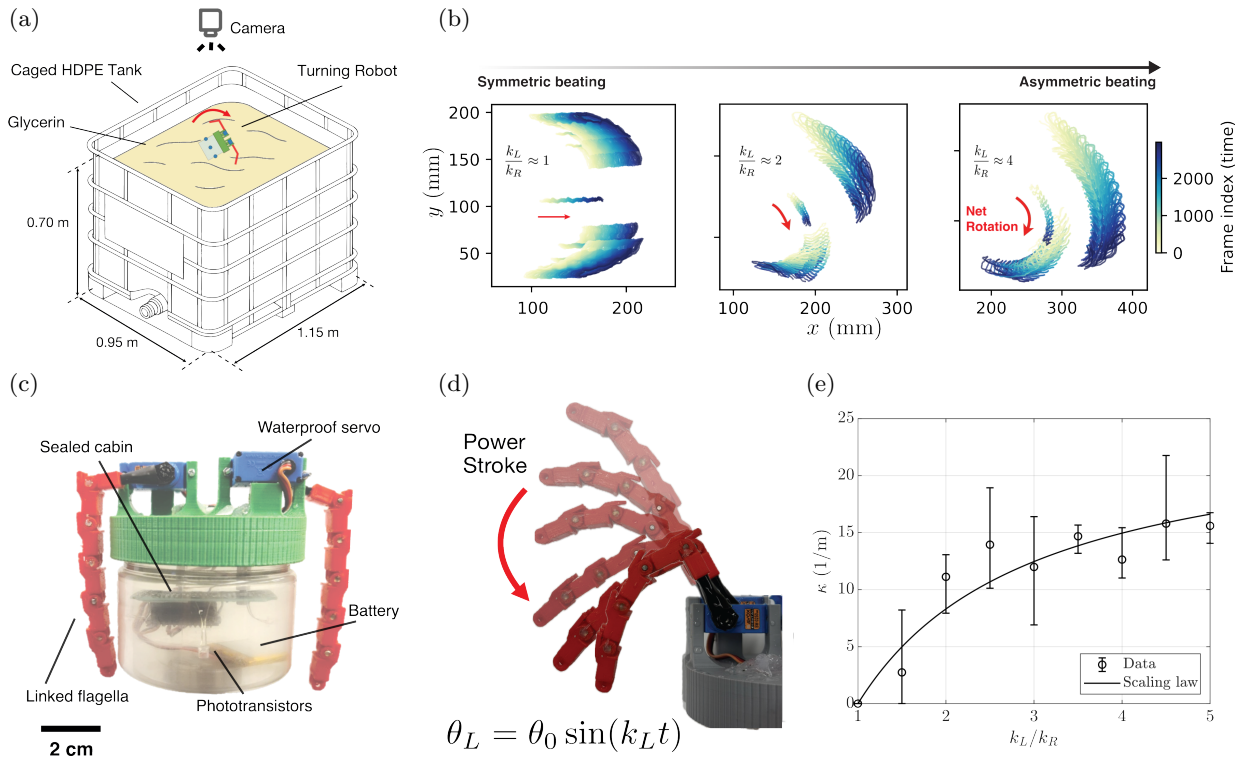

To mimic the low-Reynolds number environment experienced by biological microswimmers, experiments were conducted in a large custom-built HDPE tank (roughly \SI{1}{\metre\cubed}; \cref{fig: robot}) filled with Refined 99.7\% glycerin, MIN USP K (Ingredient Supplier). 
At \SI{25}{\celsius} this fluid yields a dynamic viscosity of approximately \SI{1}{\pascal\second} (1000x that water). 
The robot was placed fully submerged beneath the fluid surface. 
For the purposes of imaging and motion tracking, a 1080p / 60 fps camera was mounted normal to the platform. 
Nine markers were placed onto the body and appendages at strategic locations to enable automatic tracking with DeepLabCut \cite{mathis2018deeplabcut}, which yielded sub-pixel \((x_i, y_i)\) coordinates per frame. The centre of geometry (CoG) was computed each frame and smoothed with a fourth-order Savitzky–Golay filter.

\subsection*{Dependence of track curvature on beat frequency asymmetry}
The two robotic appendages are each driven independently and sinusoidally at a constant frequency. 
As each cilium is actuated by a single basal motor, the associated waveform is an emergent property of the design and cannot be readily controlled; thus, we focus exclusively on frequency asymmetry for the purposes of validation. 

The path curvature $\kappa$ increased monotonically with the frequency ratio $k_L/k_R$, where $k_L$ and $k_R$ are the actuation frequencies of the two cilia, respectively. 
In the absence of any beat asymmetry $k_L = k_R$ (both appendages performing the same stroke at the same rate but mirrored), the robot body swims forward along a nearly straight path with zero curvature. 
Rotation was observed whenever the frequencies differed.
As $k_L/k_R$ increased, the radius decreased, yielding tight counter-clockwise loops by $k_L/k_R \geq 4$.
\Cref{fig: robot}e summarises the relationship between $\kappa$ and $k_L/k_R$, while representative trajectories at $k_L/k_R=1, 2, 4$ appear in \cref{fig: robot}b. Remarkably, the experimental observations are consistent with the scaling law of \cref{eq: freq: scaling}; experimental uncertainty (3 repeats per datapoint) limits the scope of quantitative comparison.

\subsection*{Implementation of phototaxis in a robotic flagellate} 
\begin{figure}
    \centering
    \begin{overpic}[width=0.65\linewidth]{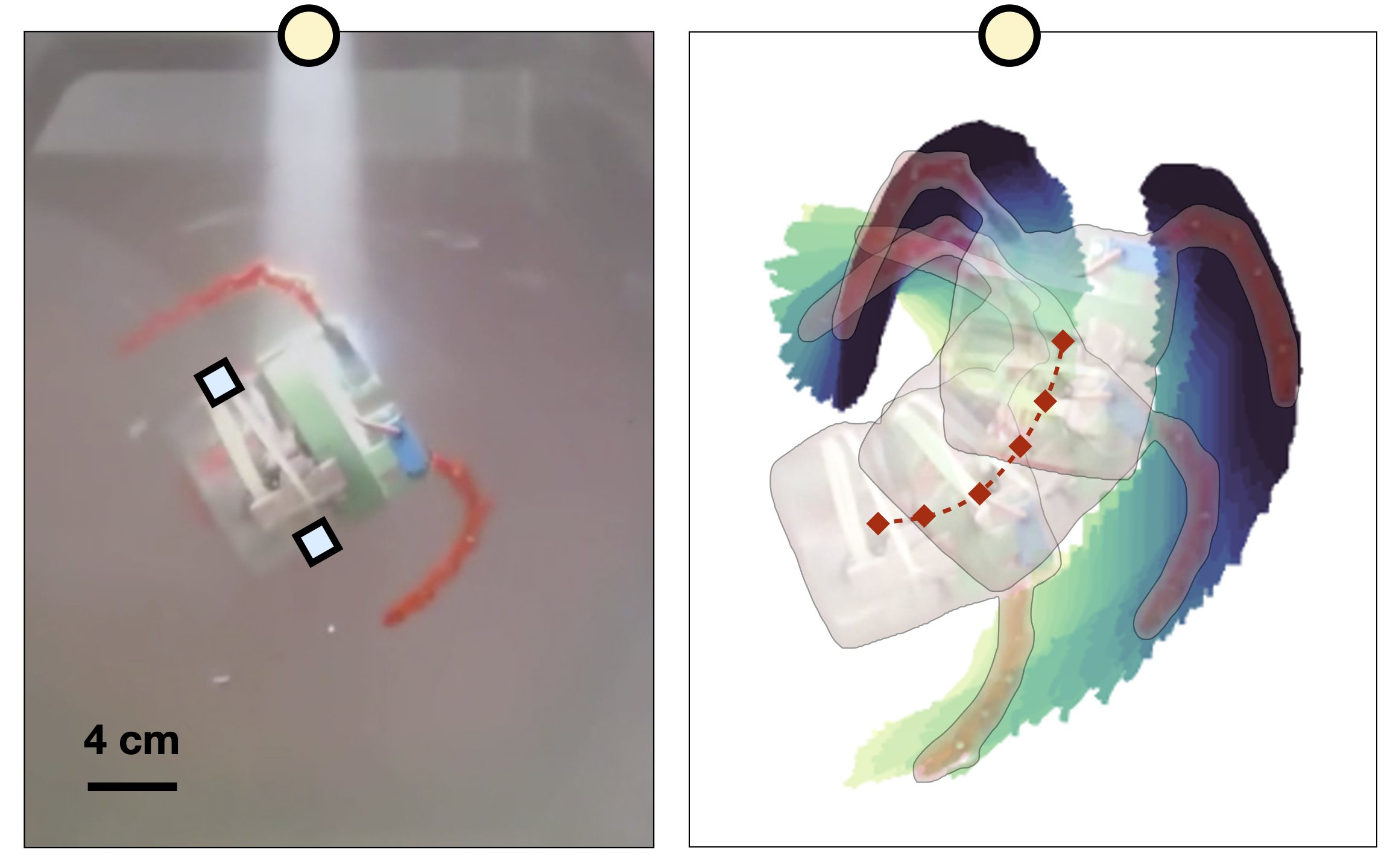}
    \put(3,54){(a)}
    \put(51,54){(b)}
    \end{overpic}
    \begin{overpic}[width=0.33\linewidth]{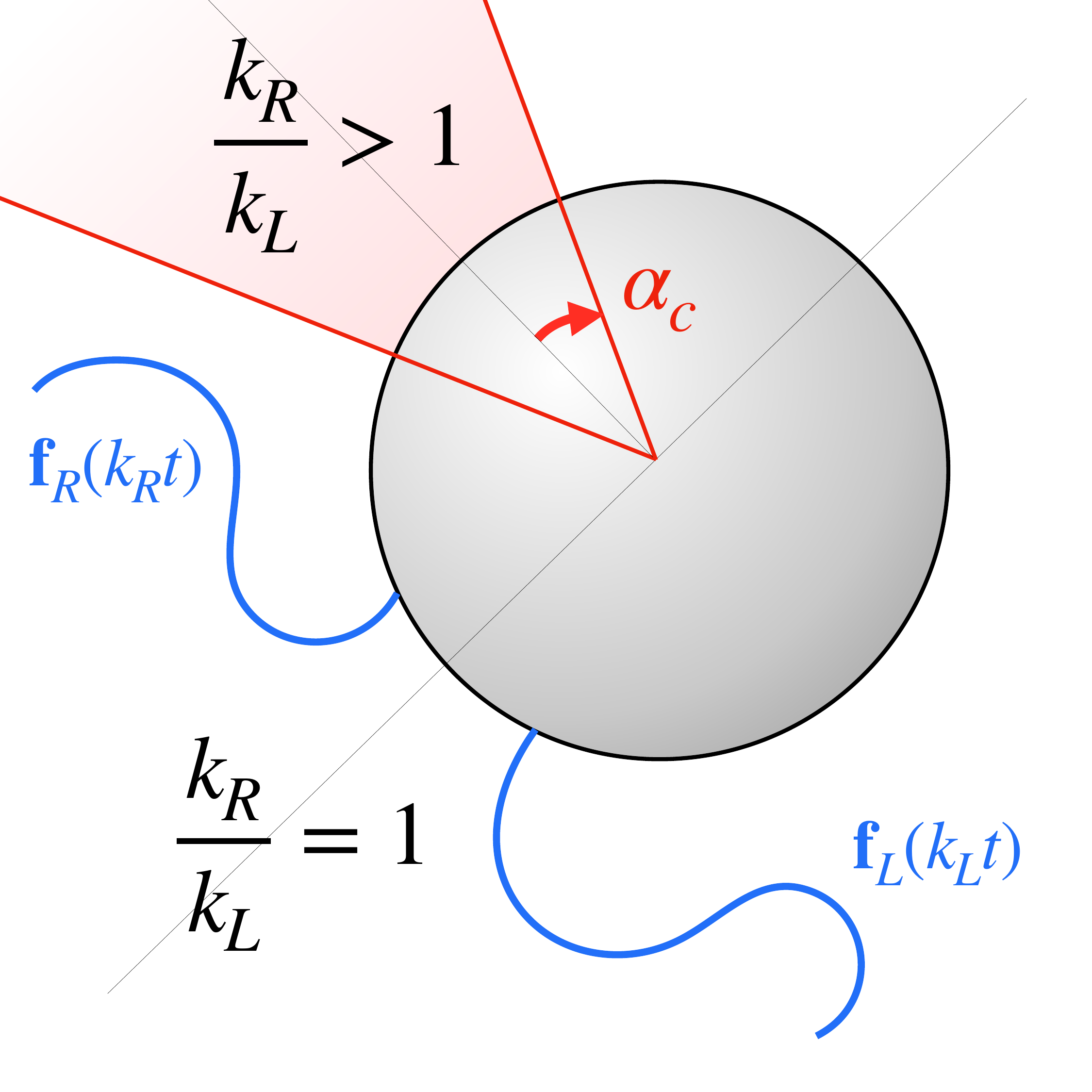}
    \put(0,106){(c)}
    \end{overpic}
    \begin{overpic}[width=0.6\linewidth]{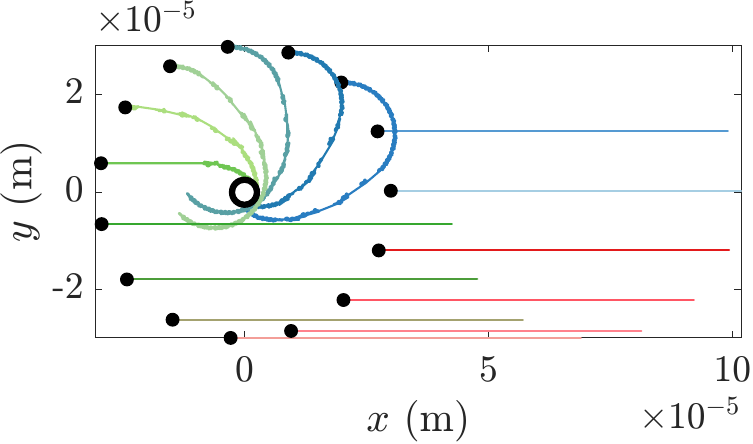}
    \put(0,55){(d)}
    \end{overpic}
    \begin{overpic}[width=0.35\linewidth]{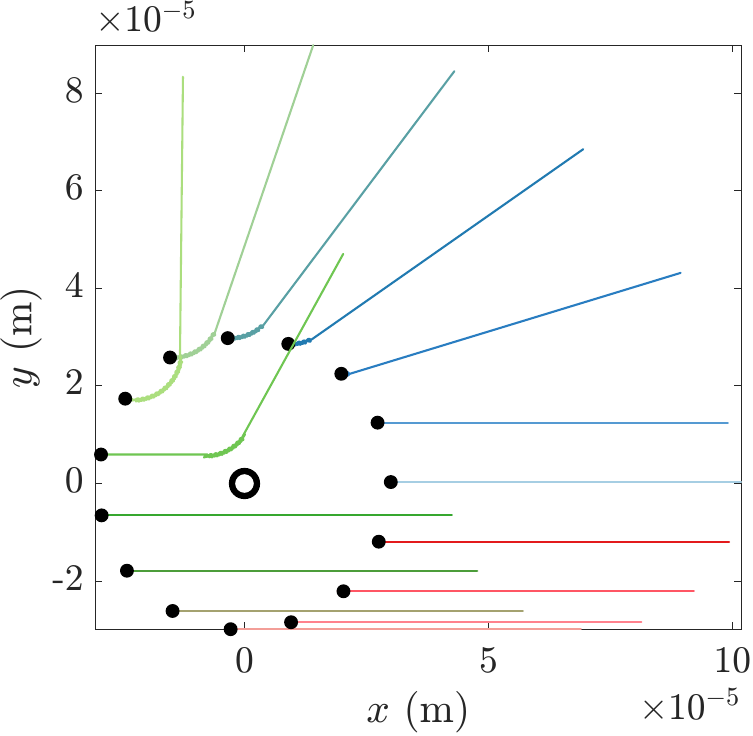}
    \put(0,96){(e)}
    \end{overpic}
    \caption{Experimental and numerical proof-of-principle that closed-loop control in a robophysical system with light sensors can lead to phototactic turning. (a) Picture of the experimental setup, showing the robotic model during phototactic motion. The robotic model is equipped with light sensors marked as blue squares. Light source position is indicated by the yellow disk. (b) Experimentally captured trajectories from the experiment illustrated in (a), overlaid with snapshots of the robot and position of its body center, showcasing positive phototaxis behaviour. (c) Principle for (negative) phototaxis in simulations using the biflagellate model. Positive phototaxis may be obtained with $k_R/k_L<1$. (d) Example trajectories for positive phototaxis, with $k_R/k_L = 0.8$ upon sensor activation. (e) Example trajectories for negative phototaxis, with $k_R/k_L = 1.5$ upon sensor activation. In (d) and (e), the light source is illustrated by a hollow black circle and the swimmer starts its trajectory at each black dot oriented at $\theta_{\mathrm{init}}$ = 0, with $\alpha_c = \pi/4$.}
    \label{fig: phototaxis}
\end{figure}

Next, inspired by the phototactic capabilities of flagellated microalgae, we sought to establish whether this turning mechanism could indeed be coupled to a sensory mechanism to achieve a stimulus-dependent response. 
We further equipped the robotic model with photosensors around the body (\cref{fig: robot}c). 
We programmed the servo motors to respond to a light-on stimulus by instantaneously changing the actuation pattern through differential frequency response in the two flagella. 
The response turns off whenever the robotic model orient away from the light and the light intensity drops below a threshold. 
In response to a stimulus from above, the robotic model is seen to turn towards the light (\cref{fig: phototaxis}a,b).  
To the best of our knowledge, this provides the first proof-of-principle demonstration of close-loop phototaxis in a robotic flagellate. 

Finally, we extend our simulated biflagellate swimmer model to replicate this phototaxis response. We adopt a simplified representation of the sensor, schematised in \Cref{fig: phototaxis}c, assuming that asymmetric swimming is triggered whenever the light source falls within the cone defined by an angular opening of $\pm\alpha_c$ relative to the sensor's orientation. 
Both negative and positive phototaxis can be obtained by setting $k_R/k_L >1$ and $k_R/k_L < 1$ upon sensor activation, respectively. Results for different initial orientations with respect to the light are presented in \cref{fig: phototaxis}d,e. The model exhibits phototactic behaviour when the light illuminates the sensor side, corroborating the experimental observations in the robophysical system.

\section*{Discussion}

In this work, we comprehensively explored the feasibility of controlling microscale swimming motility by manipulating bilateral asymmetry in a conceptually simple swimmer design featuring a symmetric body geometry, inspired by biflagellate self-propulsion.
Through a series of models of increasing complexity and realism, we formulated an approach to quantifying the degree of asymmetry, through amplitude, frequency or phase shift. In the cases of frequency and amplitude asymmetry, this yielded a universal scaling law between the resulting trajectory curvature and the asymmetry, which includes a maximum possible curvature that is set by the swimmer geometry. We validated this fundamental scaling relationship using a robophysical model that is capable of autonomous motility at low Reynolds number, highlighting the broad applicability of the derived relationships between asymmetries in propulsion and behaviour. In the case where asymmetry is captured by the ratio between the beat frequencies of the two appendages, we also identified a novel resonant behaviour at small integer or half-integer values of frequency ratio, and rationalised this behaviour and its overall limited impact on our validated predictions. 

In the broader context of cell migration, periodic body deformations are a primary mechanism of generating net movement \cite{barnhart2010bipedal,wan2023active}. 
This is true not only of cilia-driven motility but also amoeboid or cell crawling movement, particularly during chemotaxis, which often involves alternating patterns of pseudopod extension \cite{bosgraaf2009ordered,yang2011zigzag}. 
It is intriguing to ask whether steering could operate via the same basic principle across a wide variety of motile organisms.
We remark that while frequency asymmetries appear to be not only a sufficient but also an effective mechanism for turning, our archetypal biflagellate swimmer \textit{Chlamydomonas} instead combines amplitude asymmetry and phase shifts in distinct phases of movement (for example light vs dark phase). These phases depend on the relative direction of the cell's unique eyespot and the light stimulus, resulting in phototactic reorientation. 
The two Chlamydomonas flagella have distinct ontogenetic origins: the \textit{cis} flagellum (closer to the eyespot) is attached to a basal body that is assembled \textit{de novo}, while the \textit{trans} flagellum originates from the older basal body, inherited from the mother cell. 
This makes the two flagella functionally distinct, leading to a differential response to calcium increase upon detection of the light stimulus \cite{kamiya1984submicromolar,Wan2014,dutcher2020asymmetries}, which even includes stimulus-dependent beat plane modulation \cite{wang2025light}.
Meanwhile, other biflagellates that employ two morphologically distinct flagella (one long one short, or one hairy and the other smooth) for swimming \emph{do} make use of beat frequency asymmetries (as well as amplitude) for steering \cite{tran2022coordination}. 
The divergence between beat modes of subsets of flagella presumably also underlies the phototactic responses of various quadriflagellates, though this is to be confirmed experimentally. 
How this surprisingly broad spectrum of cilia-driven movement arises from nanoscale modulation of cooperative dynein motors remains to be explored in any organism, as does understanding the complex interdependence that exists between ciliary beat pattern, waveform, frequency, and even beat plane. 

The results of this study are expected to be generalisable to a range of contexts, including quantifying helical trajectories in 3D systems, the effects of fluctuating environments, or the more general integration of sensor and motor responses. We would also hope to explore the applicability of the principles of this study to other archetypal organisms and the design of novel synthetic or hybrid microswimmers \cite{zhao2019soft,xia2024biomimetic, Paramanick2025,ketzetzi2022activity,ni2017hybrid}, especially to those that warrant extension or relaxation of some of the simplifying assumptions made in this work. There is also the opportunity for broader validation, especially in next-generation robotic systems that allow for longer observations of trajectories and more robust measurements, for instance, with the potential to improve upon the already good agreement observed in this study.

Overall, by exploring, quantifying, and highlighting minimal ingredients that are sufficient for achieving directional control in low-Reynolds number swimming, this work offers practical solutions and principles for synthetic design inspired by the robust strategies of living organisms. 

\section*{Open Access Statement}
For the purpose of open access, the authors have applied a CC BY public copyright licence to any author accepted manuscript arising from this submission.

\section*{Acknowledgements}
BJW is supported by the Royal Commission for the Exhibition of 1851.
This project was also funded by the European Research Council under the European Union's Horizon 2020 research and innovation programme grant 853560 EvoMotion (K.Y.W). D.I.G. acknowledges support from the Dunn Family Professorship and the ARO MURI award W911NF-19-1-023.

\appendix

\section{Effective resistance constants}\label{app: eff resistance argument}
We can derive effective resistance constants from the detailed computational model of \cref{sec: computational model} (or indeed any similarly formulated model that captures the time dependence of resistance) in the case of frequency asymmetry, justifying the broad applicability of the minimal model of \cref{sec: minimal model}.

Suppose that the time-dependent resistance coefficients $\eta_r$ and $\eta_t$ share no non-constant Fourier modes with the forcing functions $f(k_R t)$ and $f(k_L t)$. This is trivially the case when $k_R/k_L$ is irrational, as this corresponds to $\eta_r$ and $\eta_t$ being aperiodic. Then, as this precludes any form of resonance from occurring, we have that
\begin{equation}\label{eq: cesaro}
    \int_0^T \eta_r(t,k_R,k_L)f(k_R t) \intd{t} \sim \frac{T\avg{\eta_r}}{P}\int_0^P f(k_R t)\intd{t}
\end{equation}
as $T\to\infty$, where $P$ is the period of $f(k_R t)$ and $\avg{\eta_r}$ is a constant effective resistance coefficient that is independent of the particular $f$, $k_R$ and $k_L$. 

In other words, in the long-time limit, the rotation generated by $f(k_R t)$ is proportional to the average of $f(k_R t)$. Notably, this constant of proportionality is independent of $f$ and $k_R$. Analogous reasoning applies to $\eta_r(t,k_R,k_L)f(k_L t)$. Hence, the effective angular dynamics with time-dependent resistance collapses onto that of the minimal model of \cref{eq: freq: model} with a constant effective rotational resistance $\avg{\eta_r}$. Applying a similar argument to the translational dynamics, we recover the simple model as formulated at the beginning of this study with effective resistance coefficients. Hence, the scaling law \cref{eq: freq: scaling} applies in significantly broader generality.

In the case where $\eta_r$ or $\eta_t$ share a non-trivial mode with $f(k_R t)$ or $f(k_L t)$, we observe a resonance-like phenomenon and \cref{eq: cesaro} does not apply. Hence, we do not see a precise match with our simple scaling laws in these cases. The resulting behaviours are visible in \cref{fig: computational model}b as small but non-trivial deviations from the scaling law. The effects of this resonance decrease as the period of $\eta_r$ and $\eta_t$ increases, with effects only visibly apparent for small integer and half-integer values of $k_R/k_L$.

\section{Estimating errors in the improved model}\label{app: errors in better model}
With reference to the model of \cref{sec: better model}, we estimate the magnitude of the neglected terms in the force balance equations that account for relative motion of the cilia when generating forces. Using the resistive force theory of \citet{Hancock1953,Gray1955}, the order of magnitude of these omitted terms can be estimated as
\begin{equation}\label{eq: better model: errors}
    C_T L \abs{\diff{\vec{X}}{t}}\,, \quad C_N \diff{\theta}{t}\frac{L^3}{3}\,,
\end{equation}
corresponding to the linear and angular velocities, respectively. Here, $L$ is the length of the cilia, and $C_N$ and $C_T$ are normal and tangential resistive coefficients satisfying
\begin{equation}
    C_N = \frac{4\pi\mu}{\log{\left(\frac{2L}{r}\right)} - \frac{1}{2}} = 2C_T\,,
\end{equation}
with $r$ the radius of the cilia. The relative error in neglecting the terms of \cref{eq: better model: errors} is therefore approximated by the ratios
\begin{equation}
    \mathcal{A_T}\coloneqq \frac{C_TL}{6\pi\mu a}\,, \quad \mathcal{A_R}\coloneqq \frac{C_NL^3}{24\pi\mu a^3}\,.
\end{equation}
For \textit{C. reinhardtii}, estimating $L=\SI{10}{\micro\metre}$, $a=\SI{5}{\micro\metre}$ \cite{harris2001}, and $r=\SI{0.12}{\micro\metre}$ \cite{ringo1967}, we compute
\begin{equation}
    \mathcal{A_T} \approx 0.14\,, \quad \mathcal{A_R} \approx 0.29.
\end{equation}
Hence, at least with moderate accuracy, we can neglect these terms in the case of \textit{C. reinhardtii}. We will later revisit the validity of the predictions of this model by comparison with a detailed computational model that explicitly accounts for these terms.

\section{Analytical validation of the curvature--amplitude scaling law}
\label{app: better model: amplitude}
Utilising the refined model of \cref{sec: better model}, we revisit the case of amplitude asymmetry and the curvature scaling law of \cref{eq: amp: scaling}. In this model, we encode amplitude asymmetry by taking $f_L^{(i)}$ to be of the form
\begin{equation}
    f_L^{(i)} = \alpha f_R^{(i)} = \alpha k_R[A^{(i)} + B^{(i)}J(k_R t)] \,,
\end{equation}
where $i\in\{1,2\}$, $\alpha=A_R/A_L>1$ without loss of generality, and $f_R^{(i)}$ as given in \cref{eq: better model: f_R}.

Inserting this into the model of \cref{eq: better model} and integrating the $\theta$ equation in time gives
\begin{equation}
    \theta(t) = k_R (1-\alpha) ( \tilde{\beta}_1 t + \tilde{\beta}_2 K(t))\,,
\end{equation}
assuming $\theta(0)=0$ and with $K$ the antiderivative of $J$ satisfying $K(0)=0$. The constants $\tilde{\beta}_1$ and $\tilde{\beta}_2$ are defined as 
\begin{subequations}
\begin{align}
    \tilde{\beta}_1 & = \frac{A^{(1)}\sin{\psi} + A^{(2)}\cos{\psi}}{8 \pi a^2 \mu}, \\ 
    \tilde{\beta}_2 & = \frac{B^{(1)}\sin{\psi}  + B^{(2)}\cos{\psi}}{8 \pi a^2 \mu}\,.
    \label{eq: theta amplitude}
\end{align}
\end{subequations}
If $\tilde{\beta}_2$ is sufficiently small, so that periodic fluctuations around the mean angular movement may be neglected, then $\theta$ is approximately linear in $t$, and we have 
\begin{equation}\label{eq: improved model: amplitude: rotation rate}
    \theta(t) \approx \tilde{\beta}_1 k_R (1-\alpha)t\,,
\end{equation}
i.e. the swimmer rotates at a constant rate with period $2\pi/[\tilde{\beta}_1 k_R(\alpha - 1)]$.

With $\theta$ approximated in this way, we can integrate the corresponding translational equations as in \cref{sec: better model} to give
\begin{equation}
    6\pi\mu a \tilde{\beta}_1 k_R \begin{pmatrix}
        x\\ y
    \end{pmatrix}
    =
    \begin{pmatrix}
        \sin{\theta} & \cos{\theta}\\
        -\cos{\theta} & \sin{\theta}
    \end{pmatrix}
    \begin{pmatrix}
        \frac{1+\alpha}{1-\alpha}A^{(1)}\\ A^{(2)}
    \end{pmatrix}\,,
\end{equation}
neglecting terms involving $B^{(1)}$ and $B^{(2)}$ and setting constants of integration to zero without loss of generality. This corresponds to circular motion with unsigned curvature
\begin{equation}
    \kappa = \frac{\abs{6\pi\mu a \tilde{\beta}_1 k_R}}{\sqrt{\left(\frac{1+\alpha}{1-\alpha}A^{(1)}\right)^2 + \left(A^{(2)}\right)^2}}.
\end{equation}
Hence, the model predicts constant-curvature motion under the relatively weak assumption that average propulsion dominates oscillations around that average. Analogously to the main text, if we assume that $\abs{(1-\alpha)A^{(2)}}\ll\abs{(1+\alpha)A^{(1)}}$, then this curvature relation collapses onto the form found in \cref{sec: minimal model: amp}, with $\alpha = A_R/A_L$.

\section{Prescribing ciliary beats}\label{app: alpha beta}
The functions $\alpha$ and $\beta$ introduced in \cref{sec: computational model} are given explicitly by
\begin{align}
    \alpha(t) &= \sin{(t)} - \frac{1}{2}\,,\\
    \beta(t) &= \cos{(t - 9/20)} - 1\,. 
\end{align}
These forms were chosen to agree qualitatively with waveforms typical of \textit{Chlamydomonas reinhardtii}.

\section{Phase shifts dominate small frequency differences}
\label{app: phase shift}

Following the observation that phase shifts generate curved trajectories in the biflagellate model, we seek simple analytical justification of this observation in general. In order to focus on the rotational dynamics, we consider the swimmer's body to be pinned (unable to translate but free to rotate). We also set $\beta_1 = \beta_2 = 0$, meaning that the flagella effectively act as rigid rods. Then, one can analytically derive an expression for the angular velocity of the swimmer's body of the form: 
\begin{equation}
    \dot{\theta} = \frac{2C_N \ell^2}{\sigma(\alpha_1,\alpha_2)} \big [ (2 \ell + 3 a \cos \alpha_1) \dot{\alpha}_1 - (2 \ell + 3 a \cos \alpha_2) \dot{\alpha}_2 \big],
\end{equation}
where $\sigma$ is a non-vanishing trigonometric polynomial and $\ell$ is the length of the cilia. Following our earlier convention, we note that taking $\alpha_1 \equiv \alpha_2$ would lead to cancellation and, hence, symmetric swimming. Now, assume that $\alpha_1$ is $T$-periodic, and that there exists $\phi$ such that $\alpha_2(t) = \alpha_1(t+\phi)$ for all times. A Taylor expansion of $\dot{\theta}$ at second order in $\alpha_1$ (noting that the linear terms vanish) and time integration over one time period of $\alpha_1$ yields, after removing the terms of zero average, the average angular velocity $\Omega$:
\begin{equation}
    \Omega = \int_{0}^T \dot{\theta} \intd{t}= \int_0^T C ( \alpha_2^2 \dot{\alpha}_1 + \alpha_1^2 \dot{\alpha}_2 )\intd{t},
\end{equation}
for some constant C, which further reduces, for small phase shift $\phi$, to 
\begin{equation}
    \int_{0}^T \dot{\theta} \intd{t} = \phi \int_0^T K \alpha_1^2 \ddot{\alpha}_1 \intd{t} + o(\phi),
\end{equation}
with the constant $K$ depending only on the hydrodynamic drag parameters $C_T$, $C_N$ and $C_{\mathrm{body}}$, and the geometric parameters $\ell$ and $a$:

\begin{widetext}
\begin{equation}
    K = 4 C_N \frac{9 a^4 C_{\mathrm{body}} - 12 \ell a [3a^2 (C_N - 2C_T) + 4 \ell^2C_N + 8 \ell a (C_N + C_T)]}{[-3a^3C_{\mathrm{body}} + 4 \ell C_N (4 \ell^2 + 6 \ell a + 3 a^2)]^2}.
    \label{eq: k constant 1}
\end{equation}
\end{widetext}
Assuming the relation $C_N = 2C_T$ as well as $\ell = 2a$, reminiscent of \textit{Chlamydomonas reinhardtii}, one obtains
\begin{equation}
    K = \frac{-3 C_N (3 C_{\mathrm{body}} + 32 C_N)}{(-3C_{\mathrm{body}}+52C_N)^2}.
    \label{eq: k constant 2}
\end{equation}
It is clear that $K$ may be nonzero, even when $\alpha$ and $\dot{\alpha}$ have zero average. Therefore, this analytically demonstrates that phase shifts can generate rotational velocity.

\bibliography{library}

\end{document}